\documentclass{article}
\usepackage{spconf,amsmath,graphicx}

\usepackage{enumitem}
\setlist{nosep, leftmargin=14pt}

\usepackage{mwe} 

\usepackage{amssymb}
\usepackage{caption}
\usepackage{subcaption}
\usepackage{multirow}
\usepackage{xcolor}
\usepackage{amsthm}
\usepackage{cleveref}
\usepackage{bigstrut}

\usepackage{graphicx}
\usepackage{multirow}
\usepackage{amsfonts}
\usepackage{amsmath}
\usepackage{booktabs}
\usepackage{algorithm}
\usepackage{algorithmic}
\usepackage{xcolor}
\usepackage{sidecap, caption}

\title{Modality-agnostic Style Transfer for Holistic Feature Imputation}

\name{\normalsize Seunghun Baek$^1$\sthanks{Equal contribution.} \qquad Jaeyoon Sim$^{1*}$ \qquad Mustafa Dere$^{2}$ 
\qquad Minjeong Kim$^{3}$ 
\normalsize \qquad Guorong Wu$^{2}$ \qquad Won Hwa Kim$^{1}$\vspace{-3mm}}
\address{
\normalsize $^{1}$ Pohang University of Science and Technology, Pohang, South Korea \\
\normalsize $^{2}$University of North Carolina at Chapel Hill, Chapel Hill, USA \\
\normalsize $^{3}$ University of North Carolina at Greensboro, Greensboro, USA
\vspace{-5.5mm}
}
\begin{document}

\maketitle

\begin{abstract}

Characterizing 
a preclinical stage of Alzheimer's Disease (AD) 
via single imaging is difficult as its early symptoms are quite subtle. 
Therefore, many neuroimaging studies are curated with various imaging modalities, e.g., MRI and PET, 
however, it is often challenging to acquire all of them from all subjects 
and missing data become inevitable. 
In this regards, in this paper, we propose a framework that generates unobserved imaging measures 
for specific subjects using their existing measures, thereby reducing the need for additional examinations.
Our framework 
transfers modality-specific style while preserving AD-specific content. 
This is done by domain adversarial training that preserves modality-agnostic but AD-specific information, 
while a generative adversarial network adds an indistinguishable modality-specific style.
Our proposed framework is evaluated on the Alzheimer's Disease Neuroimaging Initiative (ADNI) study and compared with other imputation methods in terms of generated data quality. 
Small average Cohen's $d$ $< 0.19$ between our generated measures and real ones  
suggests that the synthetic data are practically usable regardless of their modality type. 

\end{abstract}
%
%

\vspace{-5pt}
\section{Introduction}
\label{sec:introduction}
\vspace{-5pt}


Early diagnosis of an irreversible neurodegenerative disease such as Alzheimer's Disease (AD) 
is critical to delay its progression. 
Images are often used to observe structures and functions of the brain, 
however, the diagnosis with a single imaging modality~\cite{baek2023, sim2024} is challenging 
as the changes in preclinical stages such as Mild Cognitive Impairment (MCI) are very subtle. 
Using multiple modalities will definitely help improve confidence in the diagnosis, as the imaging scans are based on different underlying mechanisms and reflect different aspects of the disease 
such as brain structure (e.g., cortical thickness), metabolism (e.g., FDG) and protein accumulation (e.g., Tau and $\beta$-amyloid).

Despite the usefulness of adopting multiple modalities, 
it is practically infeasible as 
obtaining various imaging scans is time-consuming, costly, and burdening to each subject. 
This can be a significant barrier for patients who are potentially at risk of developing AD from MCI, 
as the high cost of imaging will discourage them from medical attention.
In practice, affordable magnetic resonance imaging (MRI) is first taken and, 
expensive Positron Emission Tomography (PET) scans with various tracers are recommended 
where MCI-specific changes are differently characterized by accumulation of proteins. 
In such a scenario, if we can accurately impute measures of unobserved modalities from existing ones, 
e.g., generating expensive fluorodeoxyglucose (FDG) measures from inexpensive MRI,  
then it would help better analysis of a patient without going through burdening imaging protocols at a significantly reduced cost. 


Recent studies revealing high correlations between different imaging modalities for AD analyses demonstrate that the scenario above is potentially feasible \cite{correlation1,correlation2,correlation3,correlation4}. 
Several works~\cite{MRI2PET1,MRI2PET2} focused on this correlation and tried to generate PET scans from MRI scans by directly applying basic generative adversarial network (GAN) or cycleGAN~\cite{cycleGAN}. 
Other models such as conditional GAN (cGAN)~\cite{cGAN} and Wasserstein GAN (WGAN)~\cite{WGAN} can be an option, however, it is difficult to naively utilize them as the number of required generators will increase as a combination of the number of modalities. 
Traditional approaches such as Mean imputation \cite{donders2006gentle} are still a golden standard even though they are unrealistic. 

Although the progression of AD from MCI is differently characterized 
by different modalities and radiotracers, 
they should commonly contain AD-specific information at AD-specific regions of interest (ROI) which we consider as ``content''. 
We focus on this modality invariant content; 
we design an architecture that first projects a sample to a modality agnostic latent space, and the modality-agnostic embedding is put on with different realistic styles of various imaging modalities.
With the proposed model, 
our goal is to universally generate 
various imaging measures for each subject that accurately reflects both the modality and AD-related information to impute missing data, without exhaustively training one-to-one mapping between different modalities.

\begin{figure}[t!]
\centering
    \includegraphics[width=0.97\linewidth]{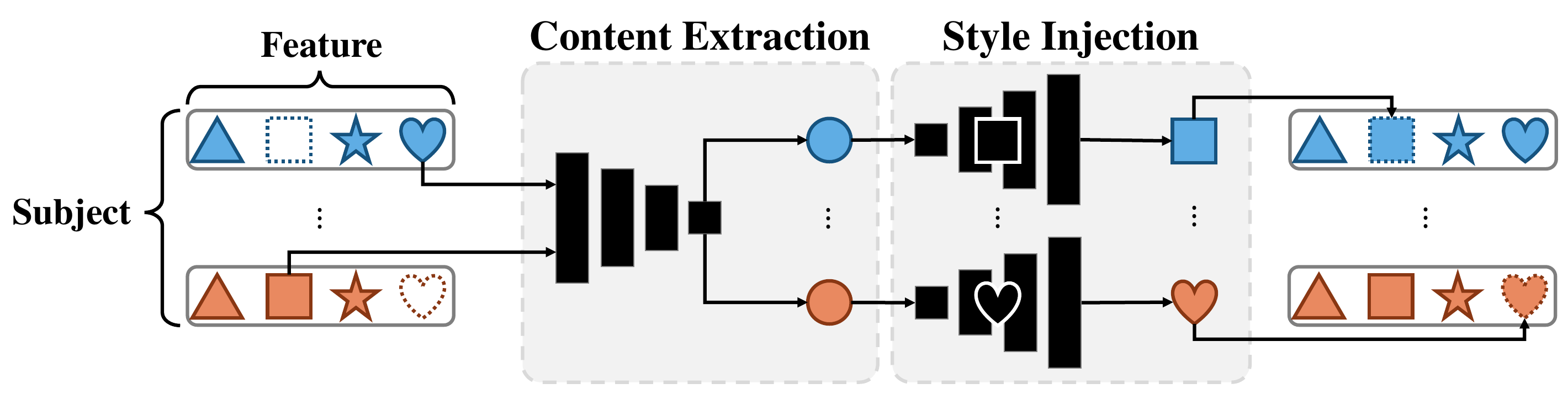}
    \vspace{-10pt}
    \caption{\footnotesize An overview of our framework. 
    In Content Extraction,
    modality-agnostic embedding is extracted from any type of feature for the subject.
    In Style Injection, 
    modality-specific generators can generate missing features not present in the original subject.
    Shape: imaging scan (i.e., domain), Color: AD-stage label (i.e.,class).
    }
    \label{fig:model_overview}
    \vspace{-17pt}
\end{figure}

\begin{figure*}[t!]
    \centering
    \scalebox{1.00}{
    \begin{tabular}{c}
        \includegraphics[width=\textwidth]{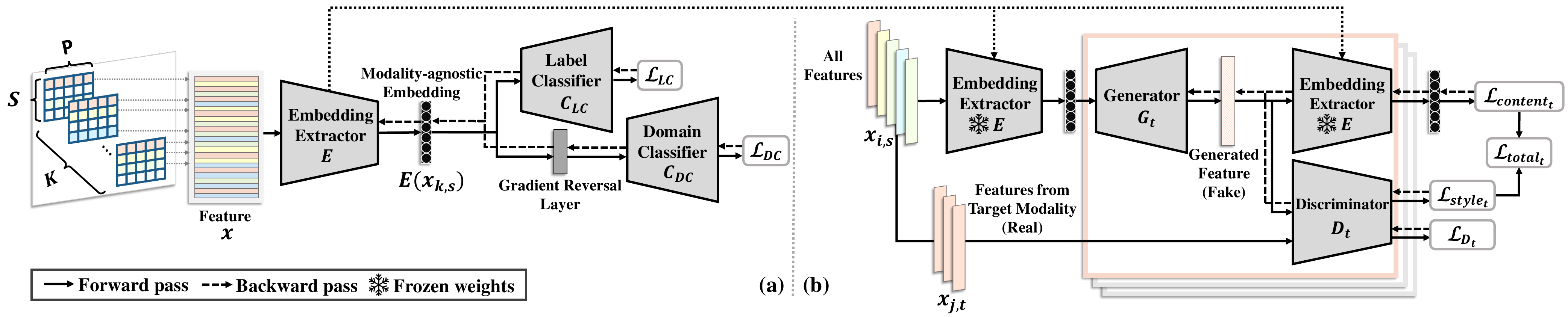}\\
    \end{tabular}}
    
    \vspace{-15pt}
    \caption{\footnotesize Illustration of our framework. (a) Content extraction, (b) Style injection.
    In (a), our framework trains embedding extractor to obtain domain-agnostic embedding which still contain label-specific information. 
    In (b), our framework endeavors to train generators from extracted domain-agnostic embedding to generate realistic measures. 
    }
    \label{fig:model_architecture}
    \vspace{-15pt}
\end{figure*}

\textbf{Key Contributions:} 
{\bf 1)} Our method generates probable estimation of unobserved imaging measures for specific subjects using their existing measures.
{\bf 2)} To the best of our knowledge, our method is {\em the first} to address numerical data imputation via style transfer. 
{\bf 3)} Experimental results on ADNI data 
demonstrate that the data generated by our method can offer sufficiently realistic information 
for downstream analyses. 




\vspace{-10pt}
\section{Method}
\label{sec:method}
\vspace{-5pt}

Consider a dataset 
where each sample is expressed as multi-variate measures taken from multiple imaging scans. 
If a subject skips some scans, which is common in many cohorts, then the sample 
ends up with missing data that discredits downstream analyses.  
To fully utilize such data, in this section, 
we propose a novel framework that completes the measures from missing modalities by 
imputing those 
for each subject based on their existing measurements. 
This is done by extracting modality-agnostic embedding with content only (e.g., target variable), 
and then training generators that equip realistic representations according to a specific modality. 
Our proposed framework consists of two consecutive parts: (i) modality-agnostic content extraction, and (ii) modality-wise style injection as illustrated
 in Fig.~\ref{fig:model_overview}, and their details are given below. 

\vspace{-10pt}
\subsection{Phase1 - Content Extraction}
\label{ssec:content_extraction}
\vspace{-5pt}

In this phase, our framework aims to train an embedding extractor $E$ to extract a modality-agnostic embedding from data. 
This embedding has semantic component as a `content' that is task-specific information (e.g., AD-specific biomarker) 
that is invariant to the domain-shift (e.g., modality).
For this, domain adversarial training \cite{domain_adversarial} is executed to make $E$ learn modality-agnostic latent space.  

Consider a subject going through $S$ different imaging scans, 
and $P$-dimensional measurements (from $P$ brain regions) are derived from individual scans.  
Then, the $k$-th sample is given as $x_k \in \mathbb{R}^{S \times P}$, where 
$x_{k,s} \in \mathbb{R}^{P}$ is denoted as a modality-specific data for the $k$-th subject and $y_k$ is its corresponding diagnostic label.  
If the subject skips $s$-th scan, the whole $x_{k,s}$ 
becomes missing. 
To train $E$ to extract modality-invariant disease-specific contents, 
individual $x_{k,s}$ is used to train the $E$ to become agnostic on $s$.  


The embedding $E(x_{k,s})$ is made modality-agnostic using 
AD-diagnostic label classifier $C_{LC}$ and modality classifier $C_{DC}$. 
The $E(x_{k,s_i})$ and $E(x_{k,s_j})$ should be similar 
where $s_i\neq s_j$ and $s_i,s_j \in \{1,2,\cdots, S\}$ for a specific subject $k$.
Concurrently, $C_{LC}$ is expected to accurately predict class labels, whereas $C_{DC}$ should exhibit uncertainty in distinguishing the type of imaging scans.
Thus, 
we use two loss functions $\mathcal{L}_{LC}$ and $\mathcal{L}_{DC}$ corresponding to $C_{LC}$ and $C_{DC}$ respectively: 
\begin{equation}
\footnotesize
\begin{aligned}
    \mathcal{L}_{LC} = \mathcal{J}(y_k, C_{LC}(E(x_{k,s}))), ~~
    \mathcal{L}_{DC} = \mathcal{J}(s, C_{DC}(E(x_{k,s})))
\end{aligned}
\end{equation}
where $\mathcal{J}$ is a loss function (e.g. Cross-Entropy).
While $C_{LC}$ and $C_{DC}$ are trained from $\mathcal{L}_{LC}$ and $\mathcal{L}_{DC}$ independently, 
$E$ is updated in the direction of decreasing $\mathcal{L}_{LC}$ and increasing $\mathcal{L}_{DC}$.
We adopt the gradient reversal layer \cite{domain_adversarial} 
that converts the sign of gradient between $E$ and $C_{DC}$ to make $E$ remove the modality-specific information, 
such that 
its imaging modality cannot be distinguished by $C_{DC}$ but still accurately classifiable on its diagnostic label by $C_{LC}$.

\vspace{-15pt}
\subsection{Phase 2 - Style Injection}
\label{ssec:style_injection}
\vspace{-7pt}

In the context of style transfer~\cite{style_transfer1,style_transfer2}, the term `style' refers to modality-specific information that encompasses various aspects of data distinct from the content.
In section~\ref{ssec:content_extraction}, 
content was defined as task-specific information that 
is invariant to the modality type.
Thus, style needs to contain a modality-specific information which is shared among data from the same modality.
To combine the given content with a target style for the imputation, 
style needs to be added to the content without 
affecting the original content.
Notice that the number of pairs increases as a combination of the number of styles, i.e., modalities, if we were to naively train pair-wise generation between different modalities. 
However, as shown in Fig.~\ref{fig:model_architecture}(b), because we extract modality-agnostic representation from \ref{ssec:content_extraction}, 
we only need to train $S$ number of generator and discriminator pairs ($G_t$, $D_t$) where $t \in \{1,2, \cdots S\}$ denotes a specific modality type. 

Training a ($G_t$, $D_t$) pair requires separate inputs. 
To train $G_t$, 
all existing $x_{i,s},$ where $i\in\{1,2,\cdots,K\}$ and $s\in\{1,2,\cdots,S\}$, are fed to 
$E$ to produce 
$E(x_{i,s})$, 
which is used as a seed to generate fake sample $G_t(E(x_{i,s}))$ which 
facilitates training of $G_t$.   
To train $D_t$, all real $x_{j,t}$ are taken conditioned on a target modality $t$, 
if $j$-th sample has $t$-th feature. 
The $x_{j,t}$ and $G_t(E(x_{i,s}))$ are fed into $D_t$ for real-and-fake discrimination, 
and $G_t$ is trained to deceive $D_t$ by generating realistic data that mimics the target modality.  
At the same time, $D_t$ is trained adversarially to differentiate real and fake.
This can be expressed using the minimax loss as 
\begin{equation}
\footnotesize
    \min_{G_t}\max_{D_t}\mathbb{E}_{x\sim p_{\text{data}(x_{j,t})}}[\log{D_t(x)}] +  \mathbb{E}_{z\sim p_{\text{data}(x_{i,s})}}[1 - \log{D_t(G_t(E(z)))}].
\label{eq:gan_objective}
\end{equation}
To optimize Eq.~\eqref{eq:gan_objective},
each loss for training $D_t$ and $G_t$ are formulated as 
{\footnotesize
\begin{align}
\label{eq:gan_loss}
    \mathcal{L}_{D_t}(x_{j,t}, x_{i,s}) =& -\log{D_t(x_{j,t})} - \log{(1-D_t(G_t(E(x_{i,s}))))}, \\
    \mathcal{L}_{G_t}(x_{i,s}) =& -\log{D_t(G_t(E(x_{i,s})))}. 
\end{align}}%
To ensure that the generated measures by $G_t$ retain the content 
with a target modality style,  
$\mathcal{L}_{content}$ is defined as a $L2$-norm 
between the original embedding $E(x_{i,s})$ and 
that of the generated data $\hat{x}_{i,t}$ $=$ $G_t(E(x_{i,s}))$ given by
\begin{equation}
\footnotesize
    \mathcal{L}_{content} = \|E(x_{i,s}) - E(\hat{x}_{i,t})\|_2.
    \label{eq:content_loss}
\end{equation}
By minimizing Eq~\eqref{eq:content_loss},
$G_t$ is guided to preserve content while generating a sample of a particular modality.
Our full objective for training $G_t$ 
is 
a combination of $\mathcal{L}_{G_t}$ and $\mathcal{L}_{content}$ as
\begin{equation}
\footnotesize
    \mathcal{L}_{total} = \alpha \mathcal{L}_{G_t} + \beta \mathcal{L}_{content}
\end{equation}
where $\alpha$ and $\beta$ are user-parameters. 
With $\mathcal{L}_{total}$, each generator is enabled to incorporate modality-specific style into the given content.
\vspace{-10pt}
\subsection{Imputation Procedure}
\label{ssec:imputation_proedure}
\vspace{-5pt}
Assume that a subject $k$ has missing feature $x_{k,s_i}$. 
Our framework can generate $\hat{x}_{k,s_i}$ from an existing feature $x_{k,s_j}$ for $s_i \neq s_j$, 
utilizing the trained content extractor $E$ and style generator $G_{s_i}$.
In this way, we ensure that all $K$ subjects have every feature for $S$ modalities for full utilization of the data. 
\begin{table}[!b]
    \vspace{-7pt}
    \caption{\footnotesize Sample-size of ADNI dataset for Different Modalities.}
    \vspace{-5pt}
    \centering
    \renewcommand{\arraystretch}{0.95} 
    \renewcommand{\tabcolsep}{1.15cm}
    \scalebox{0.34}{\huge
    \begin{tabular}{c||c|c|c|c||c}
        \toprule[1.5pt]
        \multirow{3}{*}{\textbf{Label}} & \multicolumn{5}{c}{\textbf{Modality}} \\ \cline{2-6}
        &
        
        \multirow{2}{*}{\textbf{CT}} & \multirow{2}{*}{\textbf{Tau}} & \multirow{2}{*}{\textbf{FDG}} & \multirow{2}{*}{\textbf{$\beta$-Amy}} & \multirow{2}{*}{\shortstack{\textbf{Common}\\\textbf{Subjects}}} \\ 
        & & & & & \\ \hline
        \textbf{CN} & 805 & 237 & 861 & 735 & 97 \\ \hline
        \textbf{EMCI} & 486 & 186 & 597 & 833 & 87 \\ \hline
        \textbf{LMCI} & 248 & 105 & 1138 & 447 & 38 \\ 
        \bottomrule[1.5pt]
    \end{tabular}}
    \label{tab:Datasets_ADNI}
\end{table}

\vspace{-5pt}
\section{Experimental results}
\label{sec:experiment}
\vspace{-5pt}

\subsection{Datasets}
\vspace{-5pt}

We evaluate our method on four imaging measures from MRI and PET scans from the ADNI study~\cite{jack2008alzheimer}. 
Each image was partitioned into 148 cortical and 12 sub-cortical regions using Destrieux atlas~\cite{destrieux2010automatic}.
For each parcellation, region-specific imaging features including Standard Uptake Value Ratio (SUVR)~\cite{thie2004understanding} of metabolic activity from FDG-PET, 
$\beta$-amyloid protein from Amyloid-PET ($\beta$-Amy), 
Tau protein from Tau-PET and cortical thickness (CT) from MRI were measured.
Cerebellum~\cite{rapoport2000role} was used as the reference region to standardize the SUVR values across the brain. 
Table \ref{tab:Datasets_ADNI} presents the demographics for each measure, including subjects who have taken all the imaging scans. 
A total of $N$=222 subjects have all scanned data from MRI to PET, which served as the baseline.
The diagnostic labels for each subject include cognitive normal (CN), early mild cognitive impairment (EMCI), and late mild cognitive impairment (LMCI), which we used to design the 3-way classification.

\begin{table}[!b]
\vspace{-13pt}
\caption{\footnotesize The average 
absolute Cohen's $d$ across all the ROIs 
between actual and generated distributions. 
Lower values are better, and the values $\leq 0.2$ 
are in \textbf{bold}.
}
\vspace{-7pt}
\centering
\renewcommand{\arraystretch}{1.2}
\renewcommand{\tabcolsep}{0.11cm}
    \scalebox{0.8}{\scriptsize
    \begin{tabular}{c|c||c|c|c||c|c|c||c|c|c}
    \toprule[1.5pt]
    \multirow{2}{*}{\shortstack{\textbf{Modality}\\\textbf{(Actual)}}} &\multirow{2}{*}{\shortstack{\textbf{Modality}\\\textbf{(Generated)}}} &\multicolumn{3}{c||}{\textbf{Ours}} &\multicolumn{3}{c||}{\textbf{cGAN}~\cite{cGAN}}  &\multicolumn{3}{c}{\textbf{WGAN}~\cite{WGAN}}  \\ \cline{3-11}
    & &  \textbf{CN} & \textbf{EMCI} & \textbf{LMCI} & \textbf{CN} & \textbf{EMCI} & \textbf{LMCI} & \textbf{CN} & \textbf{EMCI} & \textbf{LMCI}\\ 
    \hline

    \multirow{4}{*}{\textbf{CT}} & \textbf{CT} & 0.238 & 0.219 & 0.239 
    & 0.458 & 0.516 & 0.496 
    & 0.414 & \textbf{0.192} & \textbf{0.162} \\ \cline{2-11}
    & \textbf{Tau} & \textbf{0.122} & \textbf{0.142} & 0.358 
    & 0.315 & 0.425 & 0.401  
    & 0.212 & \textbf{0.081} & 0.222\\ \cline{2-11}
    & \textbf{FDG} & \textbf{0.141} & 0.216 & 0.231 
    & 0.216 & 0.380 & 0.431 
    & 0.255 & 0.274 & \textbf{0.063} \\ \cline{2-11}
    & \textbf{$\beta$-Amy} & \textbf{0.105} & \textbf{0.135} & 0.211 
    & 0.400 & 0.296 & 0.460  
    & \textbf{0.108} & \textbf{0.105} & 0.248\\ 
    \hline
    
    \multirow{4}{*}{\textbf{Tau}} & \textbf{CT} & \textbf{0.192} & 0.245 & 0.259
    & 0.520 & 0.466 & 0.471 
    & \textbf{0.085} & 0.237 & 0.432 \\ \cline{2-11}
    & \textbf{Tau} & \textbf{0.138} & \textbf{0.129} & 0.379
    & \textbf{0.151} & \textbf{0.149} & 0.224 
    & 0.349 & \textbf{0.116} & 0.369\\ \cline{2-11}
    & \textbf{FDG} & \textbf{0.133} & 0.206 & \textbf{0.185} 
    & 0.346 & 0.378 & 0.635 
    & \textbf{0.103} & \textbf{0.132} & \textbf{0.158}\\ \cline{2-11}
    & \textbf{$\beta$-Amy} & \textbf{0.088} & \textbf{0.107} & \textbf{0.121}
    & \textbf{0.194} & 0.309 & \textbf{0.173} 
    & \textbf{0.123} & 0.271 & 0.461 \\
    \hline
    
    \multirow{4}{*}{\textbf{FDG}} & \textbf{CT} & 0.237 & 0.217 & \textbf{0.167}
    & 0.572 & 0.644 & 0.512
    & \textbf{0.168} & \textbf{0.188} & \textbf{0.193}\\ \cline{2-11}
    & \textbf{Tau} & \textbf{0.158} & \textbf{0.140} & 0.447
    & 0.373 & 0.423 & 0.371 
    & \textbf{0.161} & 0.458 & 0.487 \\ \cline{2-11}
    & \textbf{FDG} & \textbf{0.142} & 0.213 & \textbf{0.146} 
    & 0.442 & 0.442 & 0.297 
    & 0.228 & 0.209 & 0.250 \\ \cline{2-11}
    & \textbf{$\beta$-Amy} & \textbf{0.110} & \textbf{0.106} & \textbf{0.167}
    & 0.265 & 0.300 & 0.415 
    & \textbf{0.161} & \textbf{0.150} & 0.411 \\ 
    \hline
    
    \multirow{4}{*}{\textbf{$\beta$-Amy}} & \textbf{CT}  & 0.209 & \textbf{0.185} & 0.288
    & 0.604 & 0.665 & 0.872
    & \textbf{0.186} & 0.486 & 0.716 \\ \cline{2-11}
    & \textbf{Tau} & \textbf{0.139} & \textbf{0.139} & 0.391
    & 0.267 & 0.333 & 0.290 
    & 0.423 & 0.645 & 0.888 \\ \cline{2-11}
    & \textbf{FDG} & \textbf{0.142} & \textbf{0.189} & \textbf{0.178}
    & 0.368 & 0.496 & 0.629
    & \textbf{0.076} & \textbf{0.166} & \textbf{0.135}\\ \cline{2-11}
    & \textbf{$\beta$-Amy} & \textbf{0.152} & \textbf{0.103} & \textbf{0.137}
    & 0.372 & 0.386 & 0.402 
    & 0.249 & \textbf{0.063} & 0.292\\ \hline\hline
    \multicolumn{2}{c||}{\textbf{Mean}} & \textbf{0.152} & \textbf{0.168} & 0.244 & 0.366 & 0.413 & 0.442 & 0.206 & 0.235 & 0.342\\
    \bottomrule[1.5pt]
    \end{tabular}}
\label{tab:Cohens}
\end{table}

\vspace{-5pt}
\subsection{Experiment setup}
\vspace{-5pt}

\noindent\textbf{Baselines.}
We utilized \textbf{Mean} imputation~\cite{donders2006gentle} as a classical method,
which imputed the missing values by replacing them with the class-specific means for each imaging scan.
Prominent generative models such as \textbf{cGAN}~\cite{cGAN} and \textbf{WGAN}~\cite{WGAN} also serve as baselines. 
To apply the generative models for imputation,  
we trained generators for all pairs (i.e., 16 pairs) of imaging modalities.

\noindent\textbf{Training.} 
For $E$, $C_{LC}$, $C_{DC}$, $G_t$ and $D_t$, 
Multi-Layer Perceptron (MLP) with 5, 2, 2, 13 and 5 layers were chosen. 
We used AdamW~\cite{AdamW} optimizer with learning rate $10^{-3}$, except for $D_t$ with $10^{-5}$.
Weight decay at 0.01 was adopted for every linear layer.
In Phase 1, we set the embedding size to 256 and trained $E$, $C_{LC}$ and $C_{DC}$ with $8\times10^{3}$ epoches.
In Phase 2, each $G_t$ and $D_t$ pair was trained with $3\times10^{4}$ epoches in a 9-to-1 ratio iteratively.
In $L_{total}$, $\alpha$ and $\beta$ were 1 and 100.

\noindent\textbf{Evaluation.} 
We utilized MLP with 2, 3 and 4 layers as our backbone network, progressively getting larger, 
for the downstream classification denoted as 2-MLP, 3-MLP and 4-MLP respectively.
For training of phase 1 and 2 (for $E$, $G_t$ and $D_t$), only the samples with missing measures were used 
to avoid double dipping for a downstream task. 
In the downstream classification, 
samples with missing data using imputation methods including ours were utilized during the training.  
The experiments were performed 
with 5-fold cross validation (CV) to obtain unbiased results. 

To evaluate the performance, we employed multi-class accuracy, weighted precision and recall averaged across the CV.
Each baseline was implemented and trained 
to achieve optimal outcomes for fair comparisons.
To verify whether the generated data are realistic, 
we computed the similarity between the actual and generated measurements for each modality using Cohen's $d$~\cite{rice2005comparing}.
\vspace{-8pt}
\subsection{Quantitative Results}
\label{sec:comparision_qualtitative}
\vspace{-5pt}

\begin{figure}[t!]
    \centering
    \renewcommand{\arraystretch}{1.0}
    \renewcommand{\tabcolsep}{0.1cm}
    \small{
    \scalebox{0.58}{
    \begin{tabular}{cccccl}
    &\raisebox{1\height}[0pt][0pt]{\textbf{Cortical Thickness}}
    &\raisebox{1\height}[0pt][0pt]{\textbf{Tau}} 
    &\raisebox{1\height}[0pt][0pt]{\textbf{FDG}} 
    &\raisebox{1\height}[0pt][0pt]{\textbf{$\beta$-Amyloid}} 
    & \\
    \raisebox{2\height}[0pt][0pt]{\shortstack{\textbf{Cortical}\\\textbf{Thickness}}} &
    \includegraphics[width=0.32\linewidth]{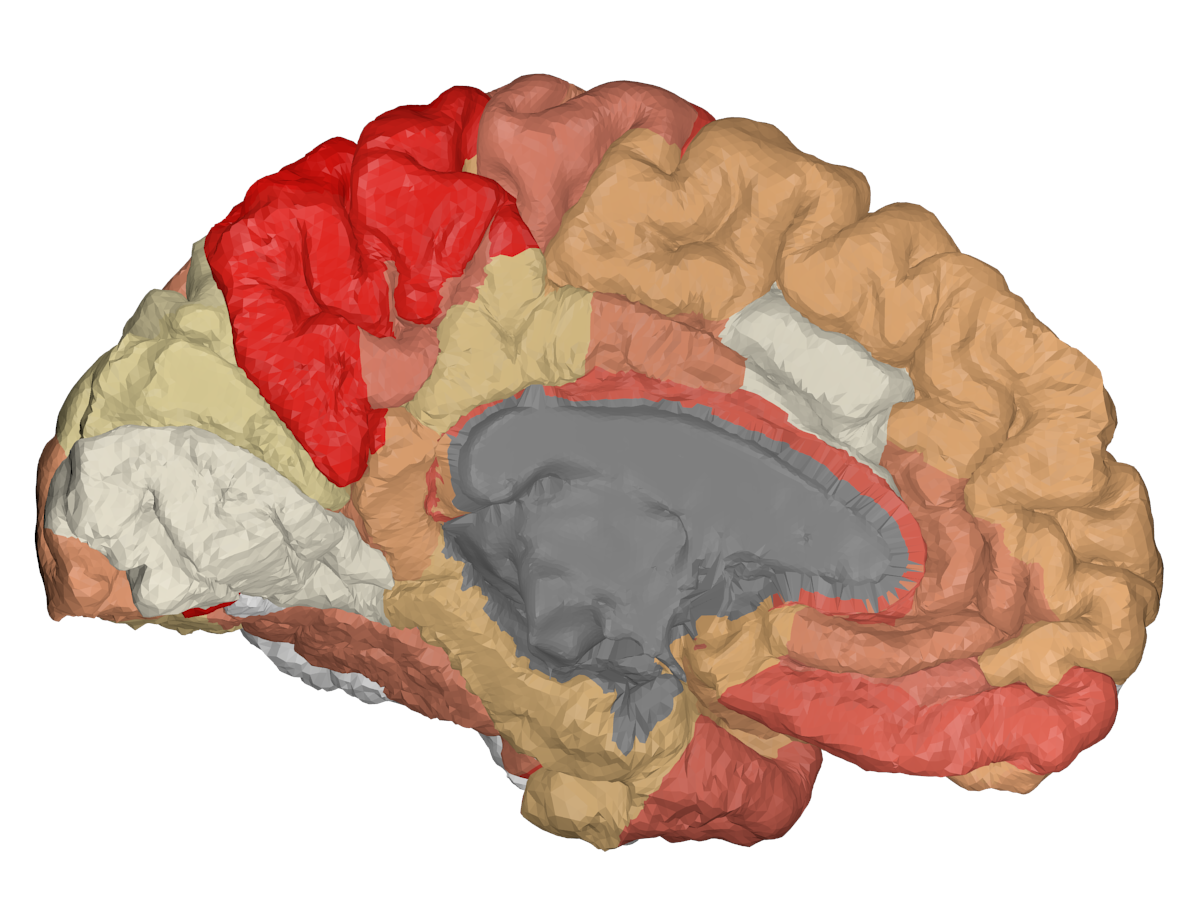}&
    \includegraphics[width=0.32\linewidth]{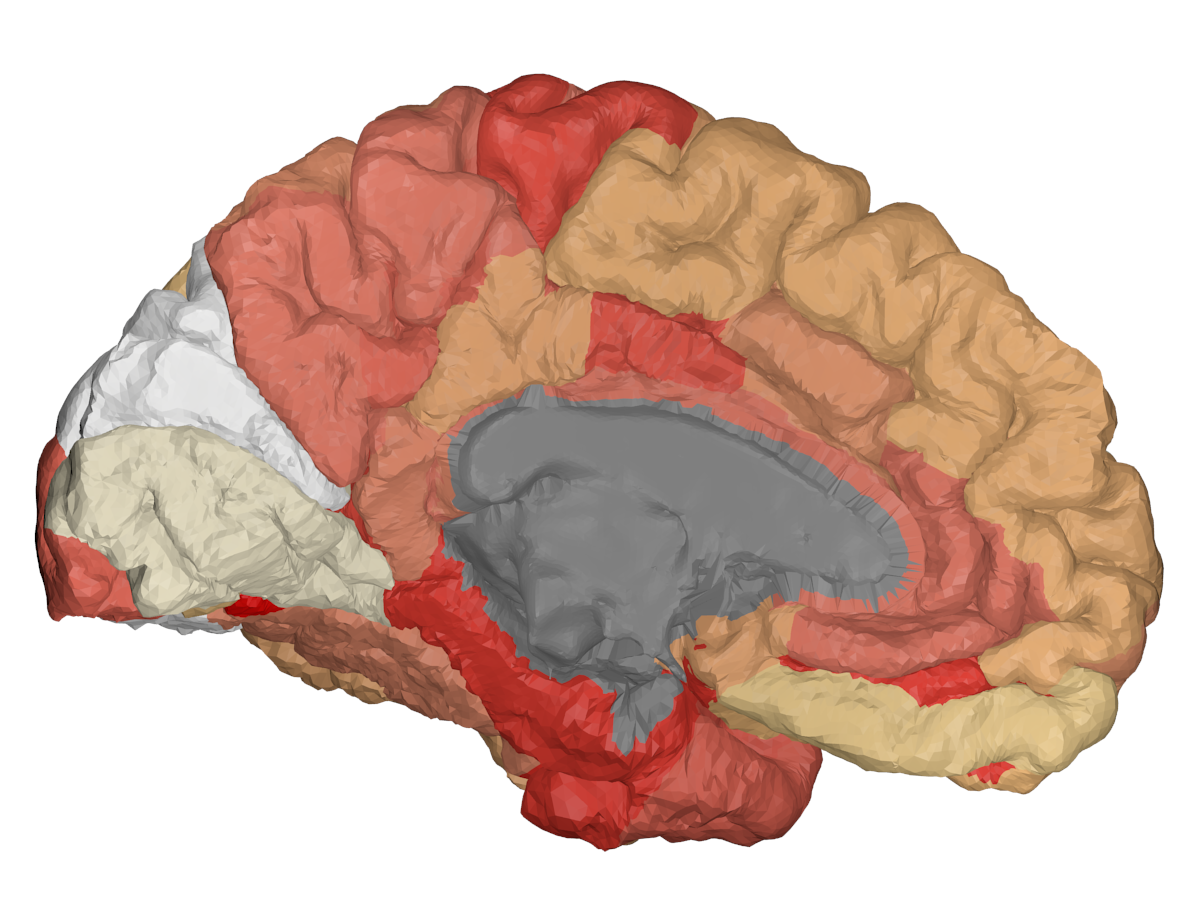} &
    \includegraphics[width=0.32\linewidth]{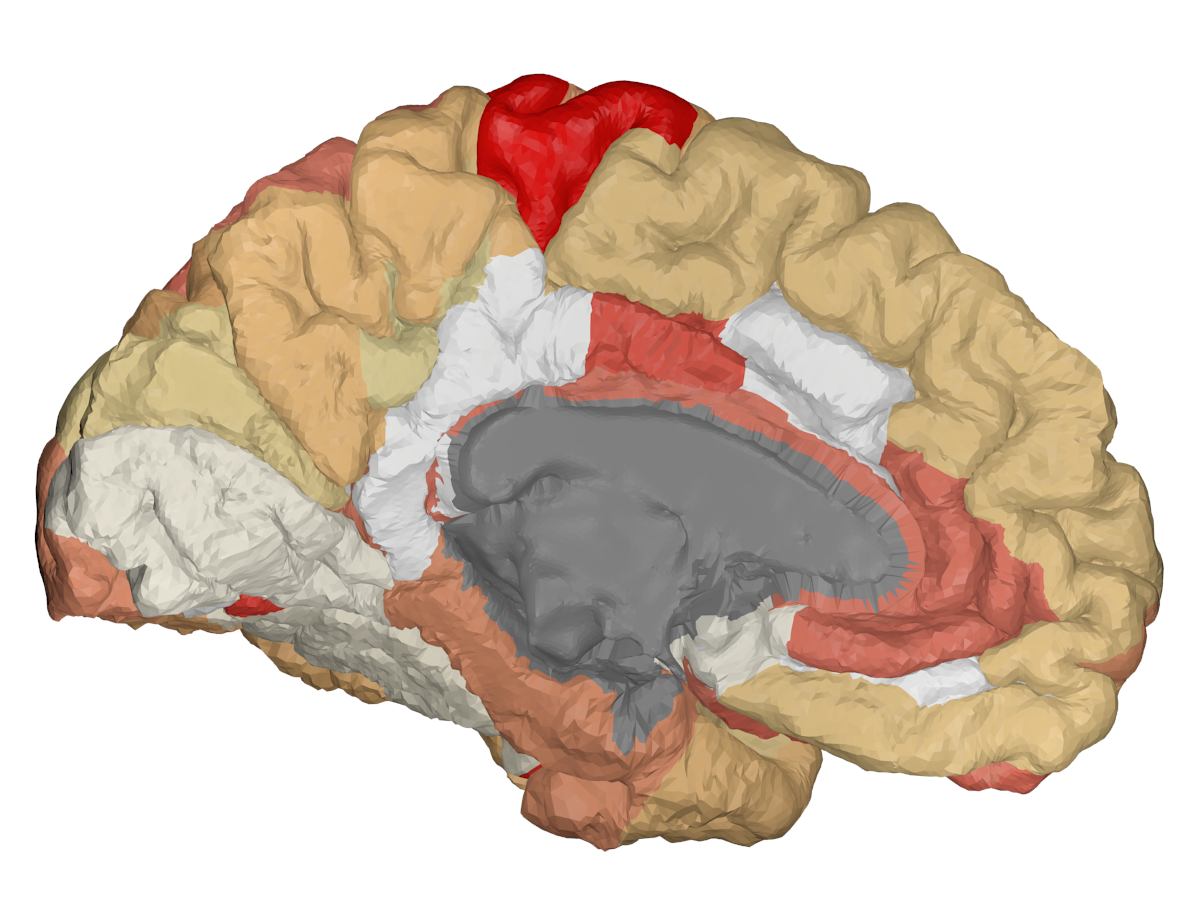} &
    \includegraphics[width=0.32\linewidth]{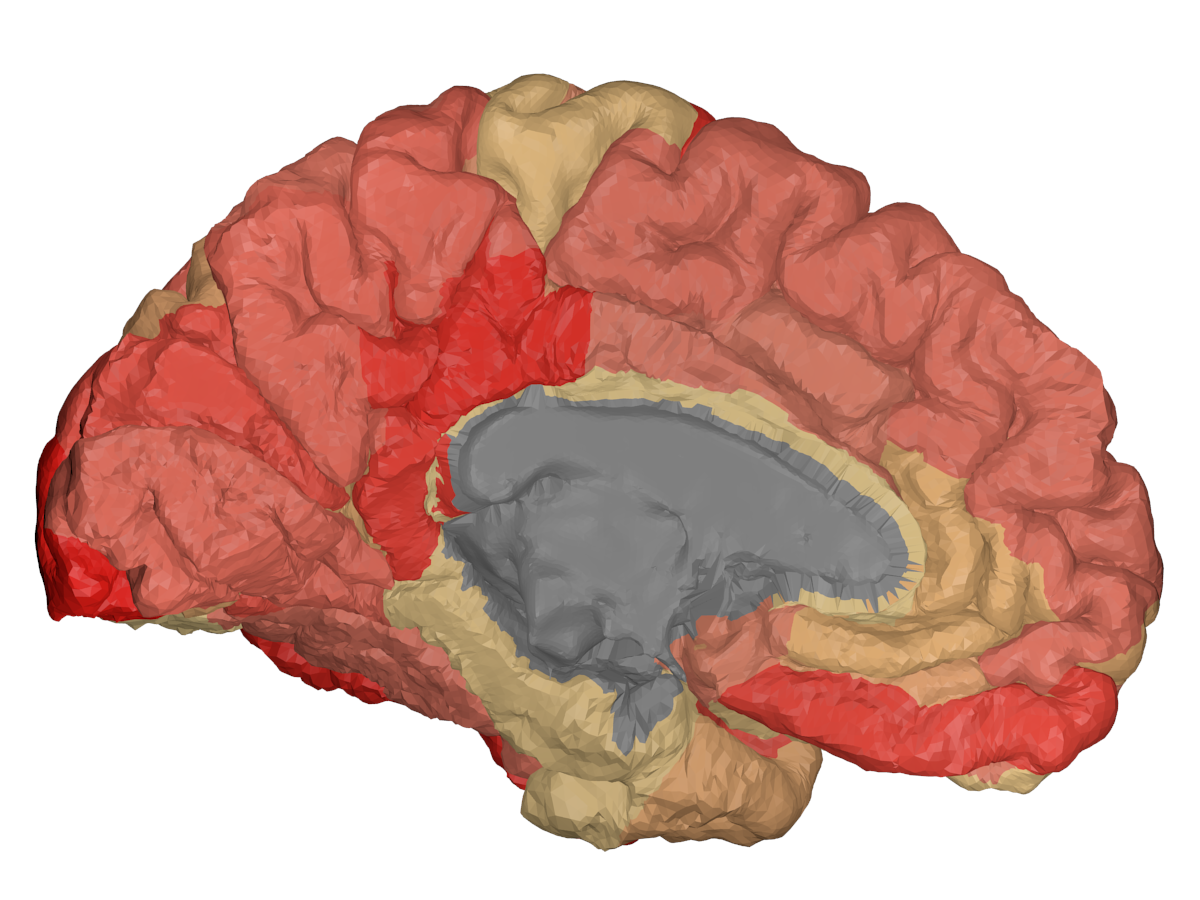} & \\
    \raisebox{5\height}[0pt][0pt]{\textbf{Tau}} &
    \includegraphics[width=0.32\linewidth]{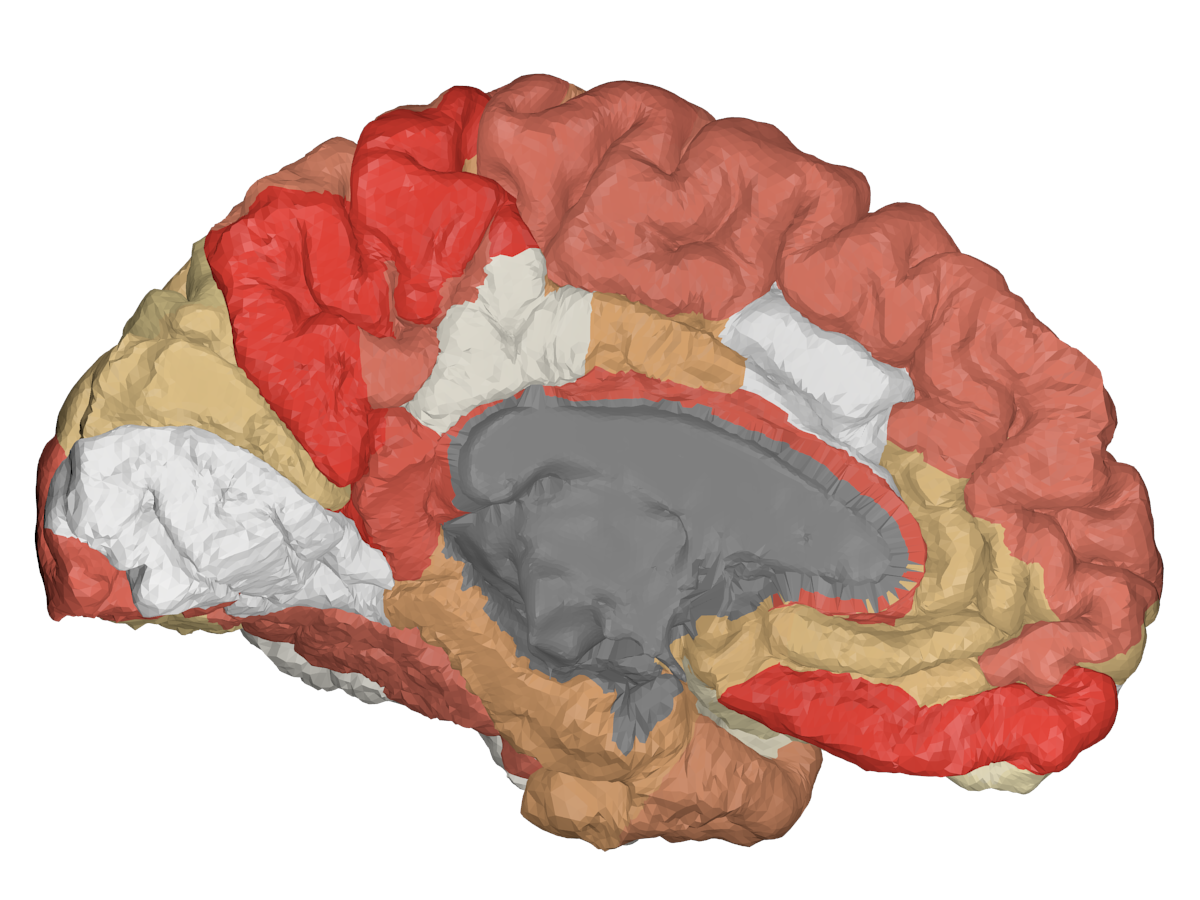}&
    \includegraphics[width=0.32\linewidth]{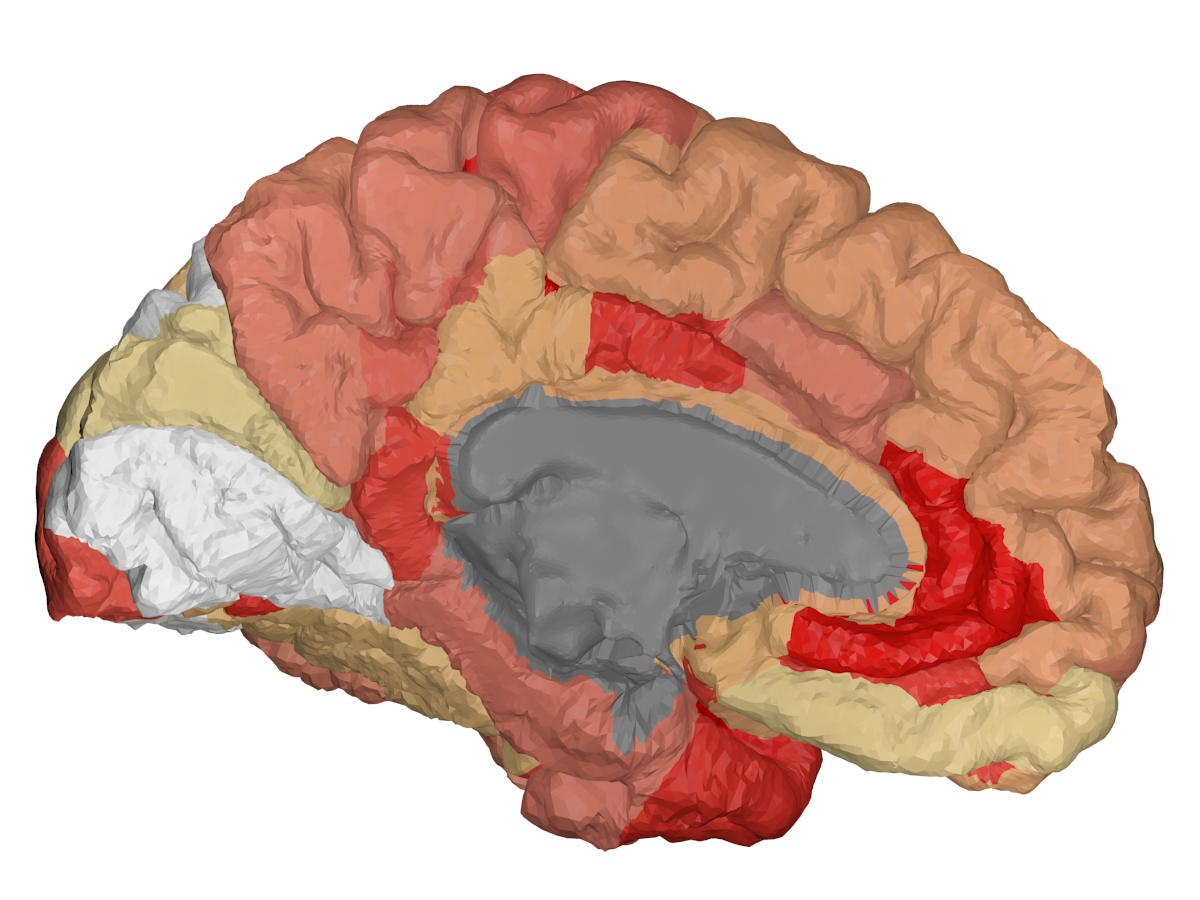} &
    \includegraphics[width=0.32\linewidth]{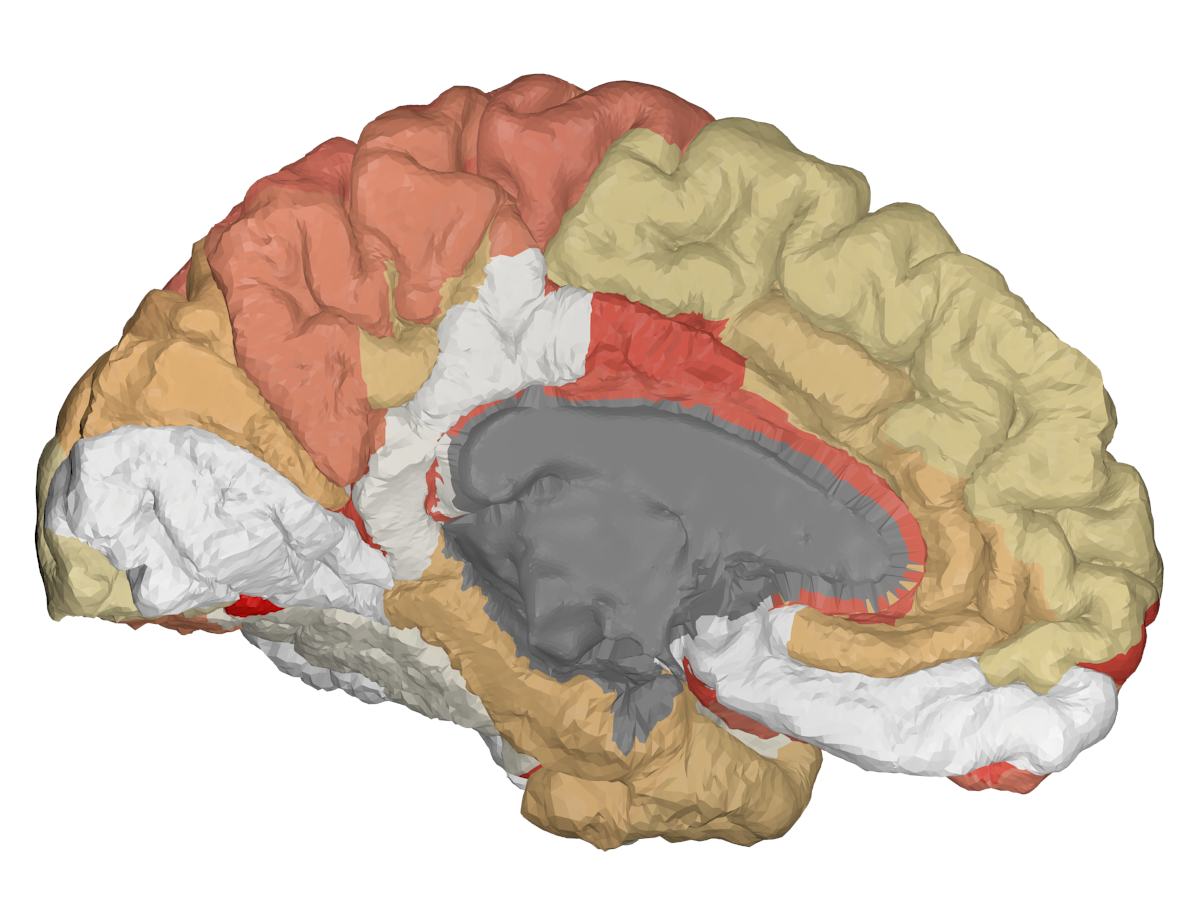} &
    \includegraphics[width=0.32\linewidth]{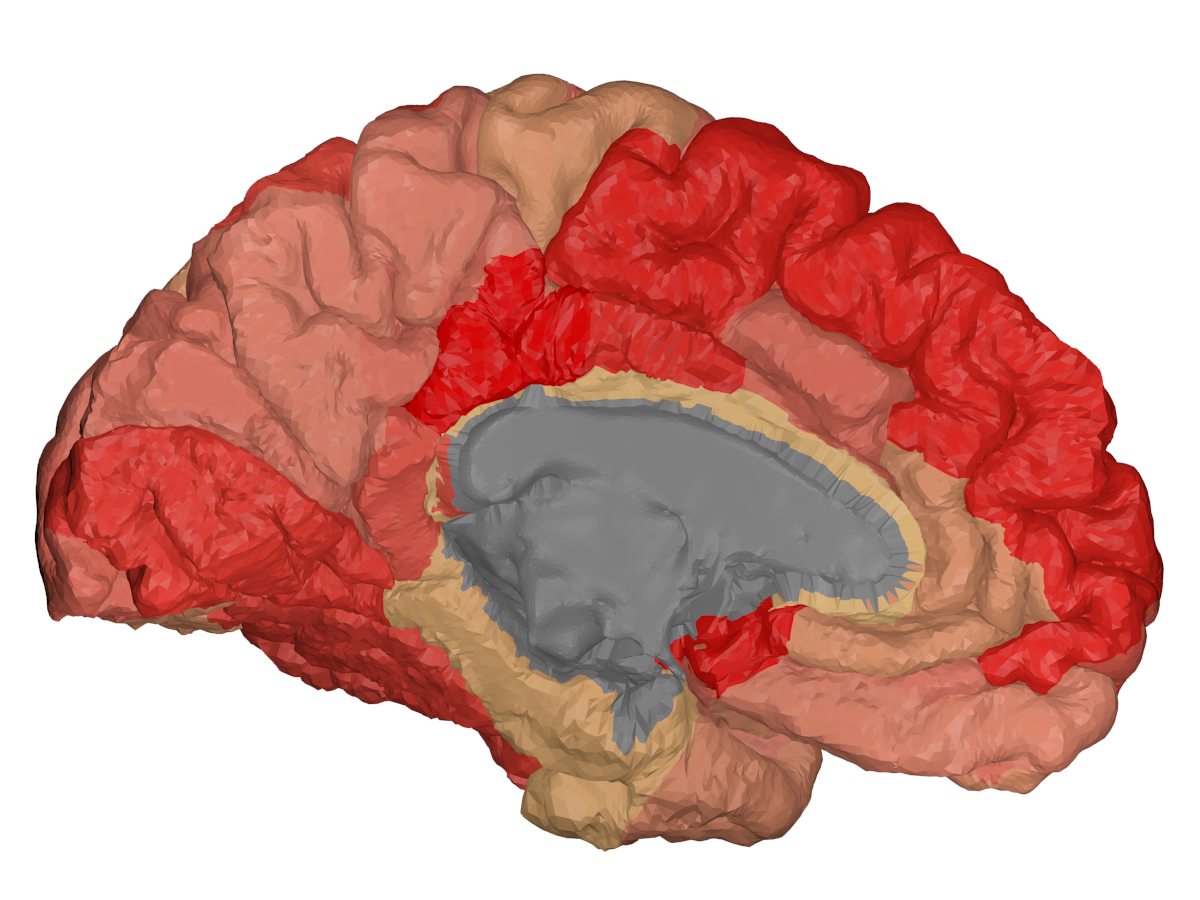} &\\
    \raisebox{5\height}[0pt][0pt]{\textbf{FDG}} &
    \includegraphics[width=0.32\linewidth]{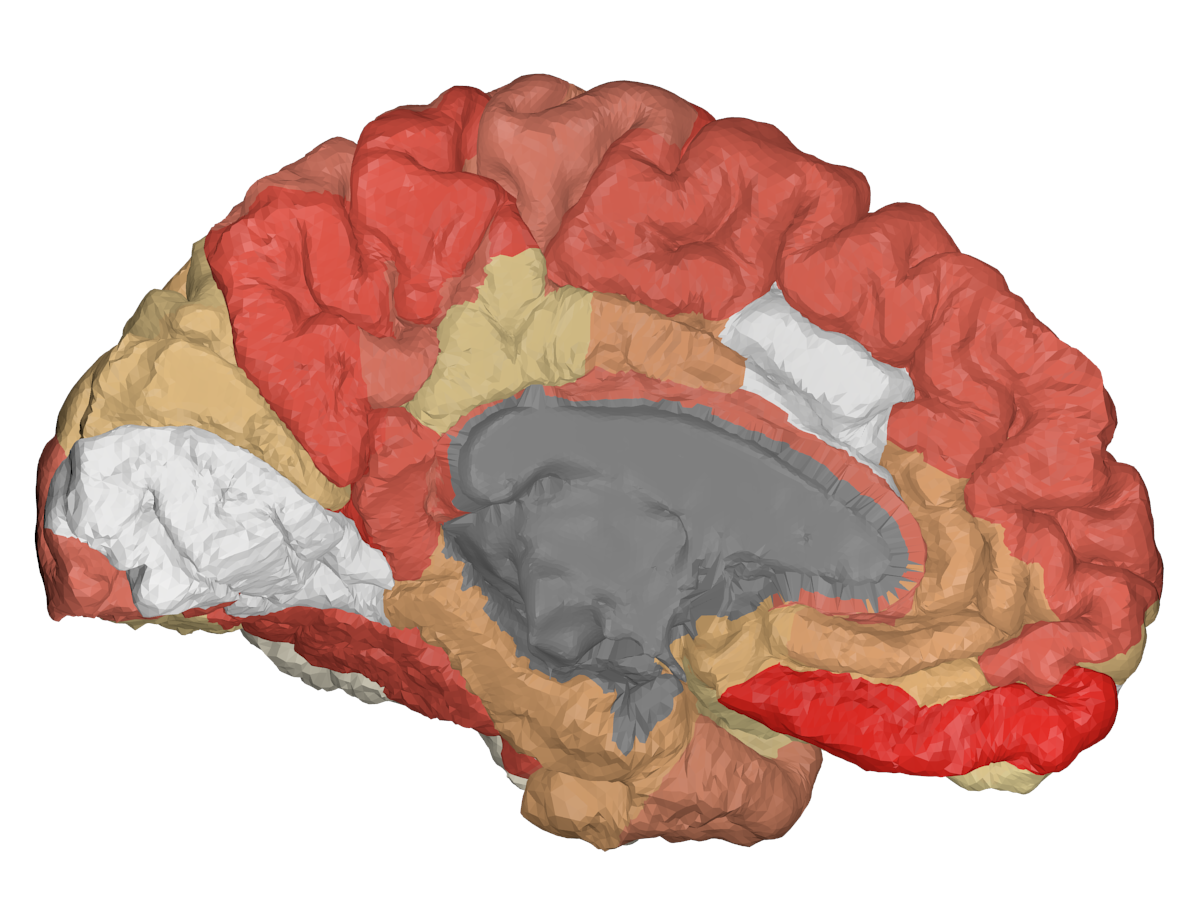}&
    \includegraphics[width=0.32\linewidth]{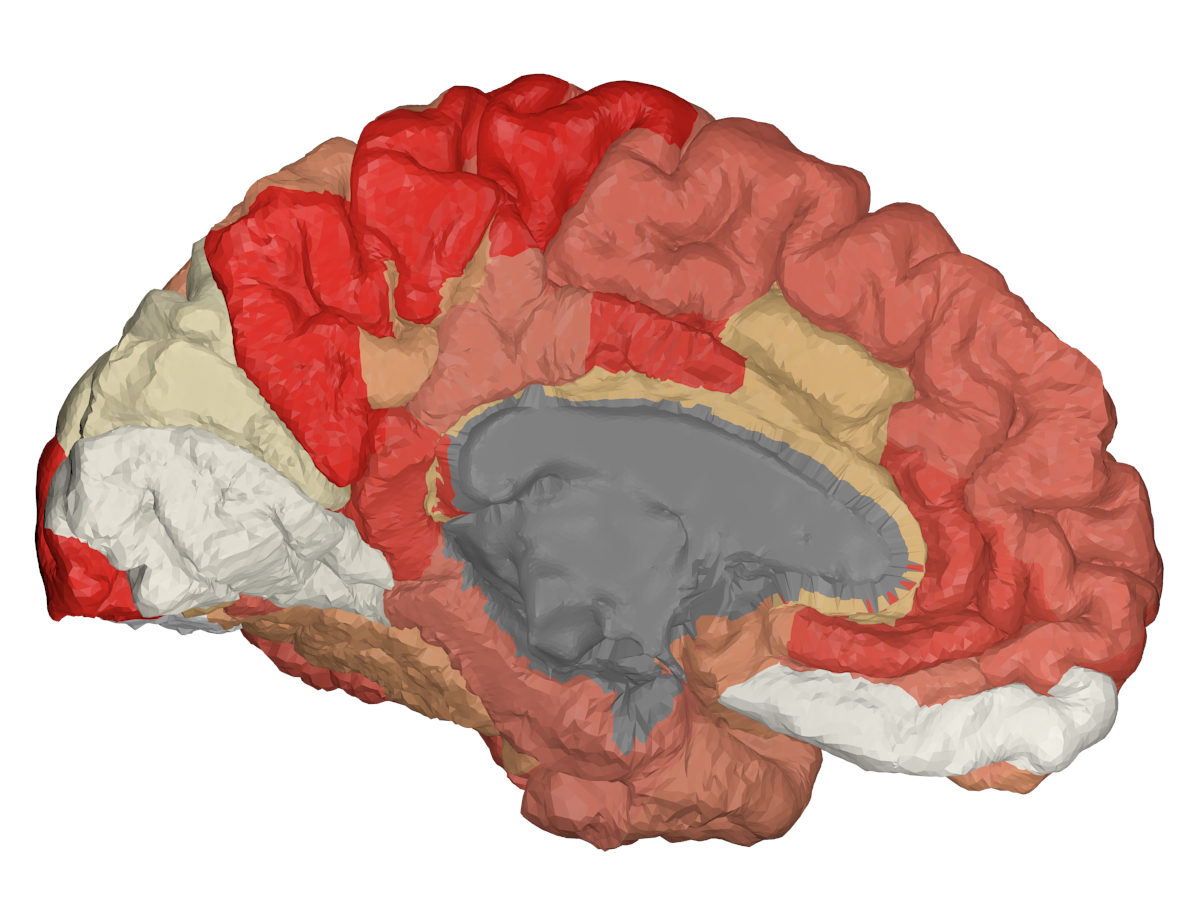} &
    \includegraphics[width=0.32\linewidth]{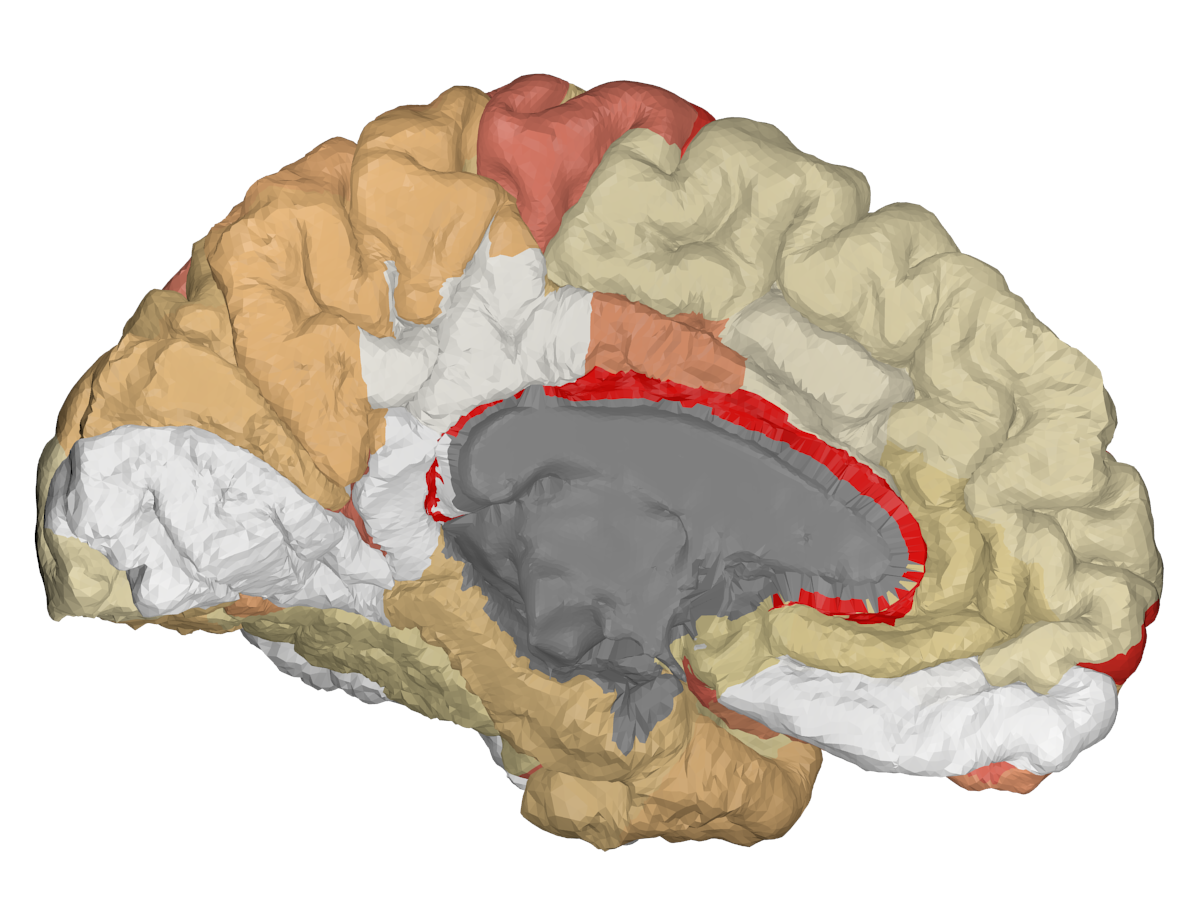} &
    \includegraphics[width=0.32\linewidth]{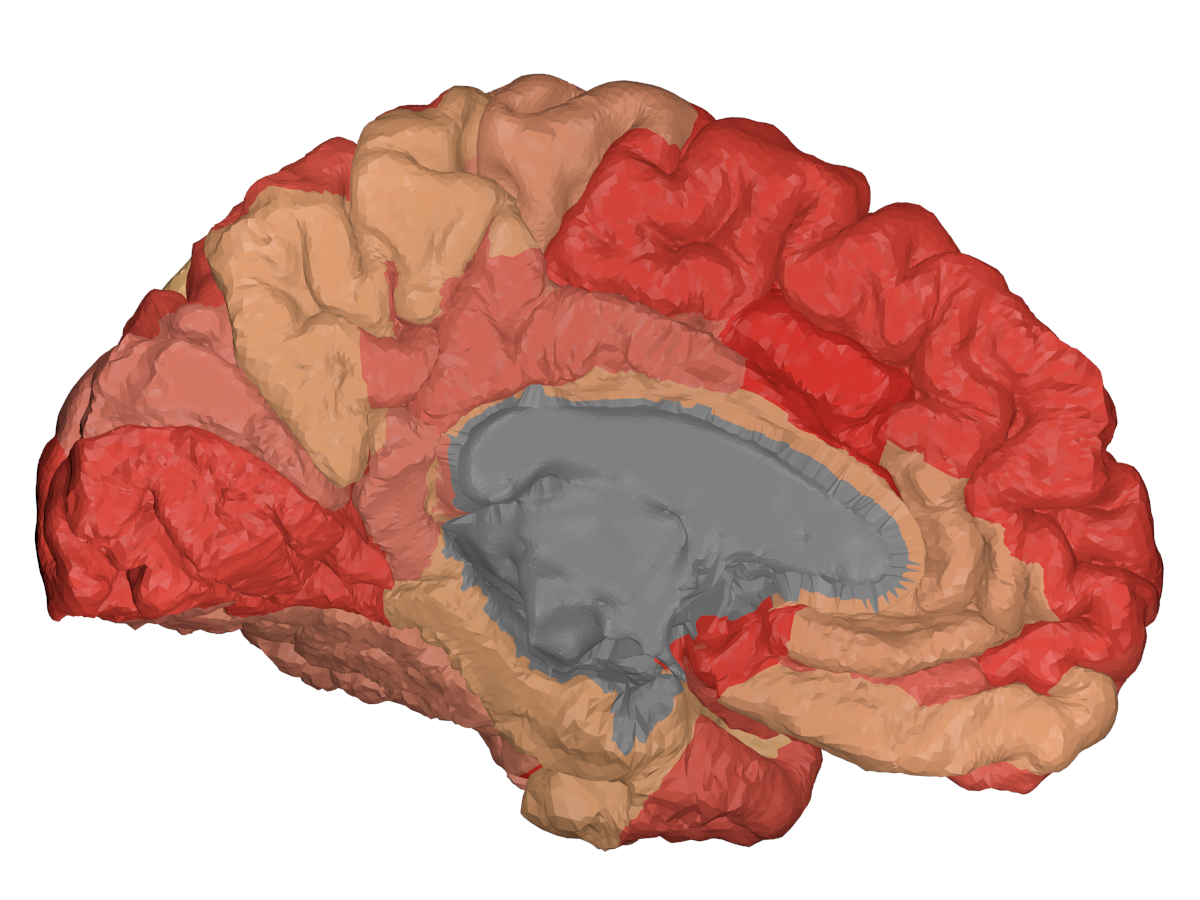} &\\
    \raisebox{5\height}[0pt][0pt]{\textbf{$\beta$-Amyloid}} &
    \includegraphics[width=0.32\linewidth]{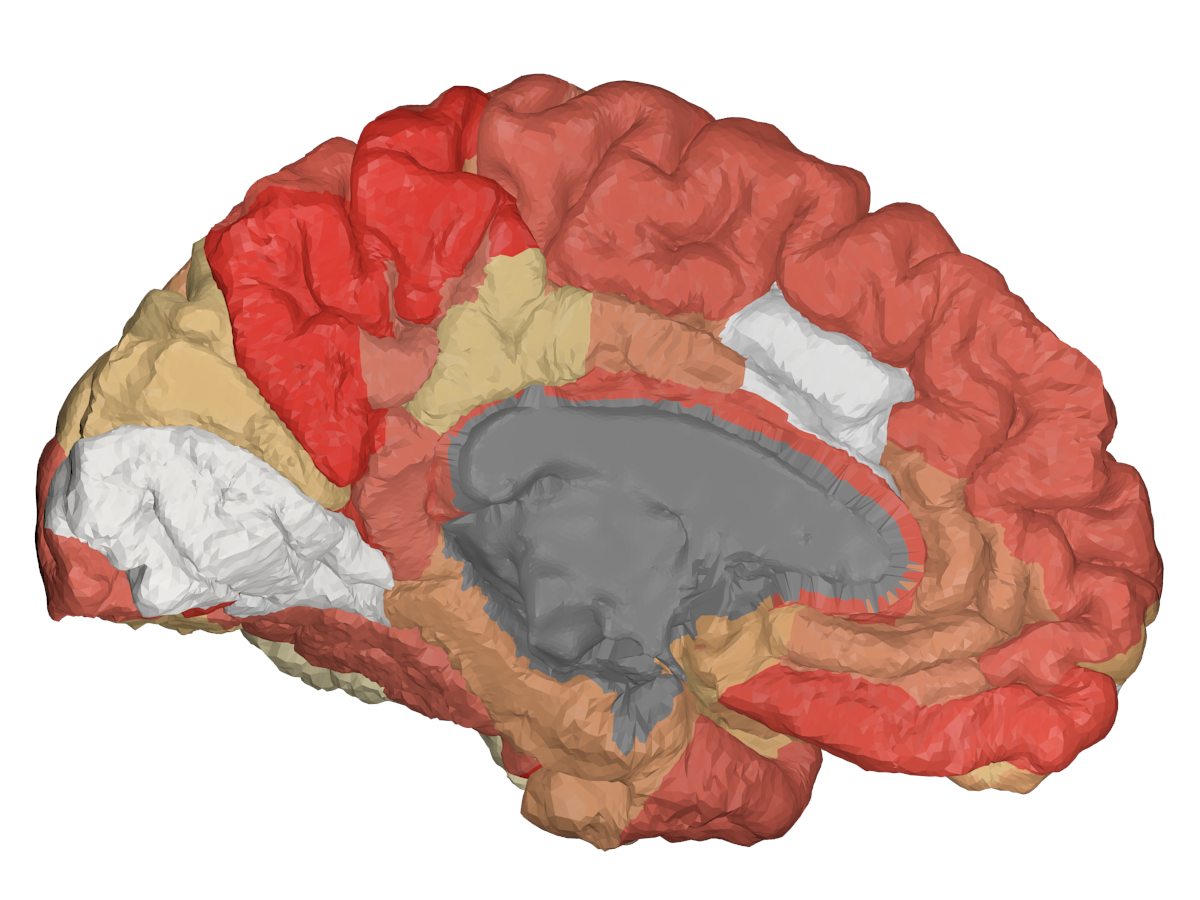}&
    \includegraphics[width=0.32\linewidth]{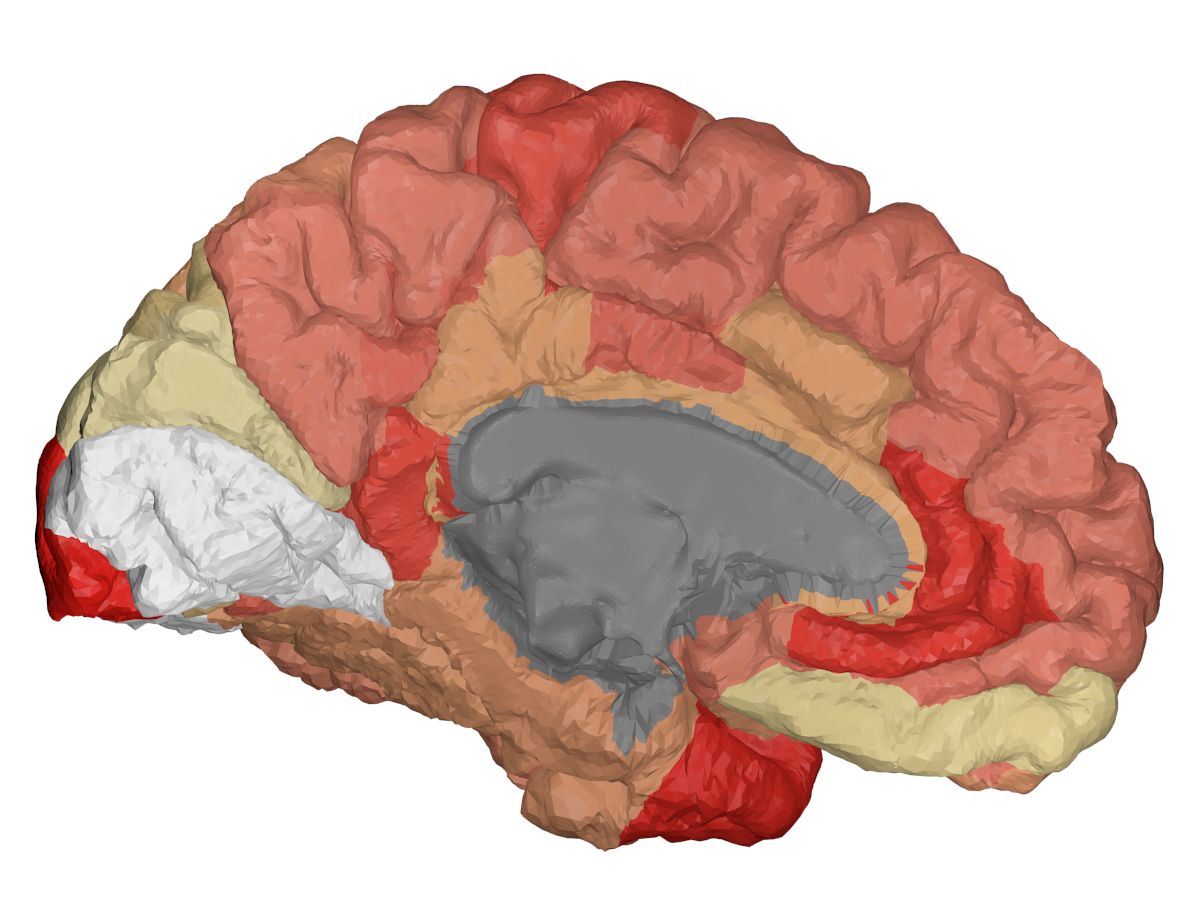} &
    \includegraphics[width=0.32\linewidth]{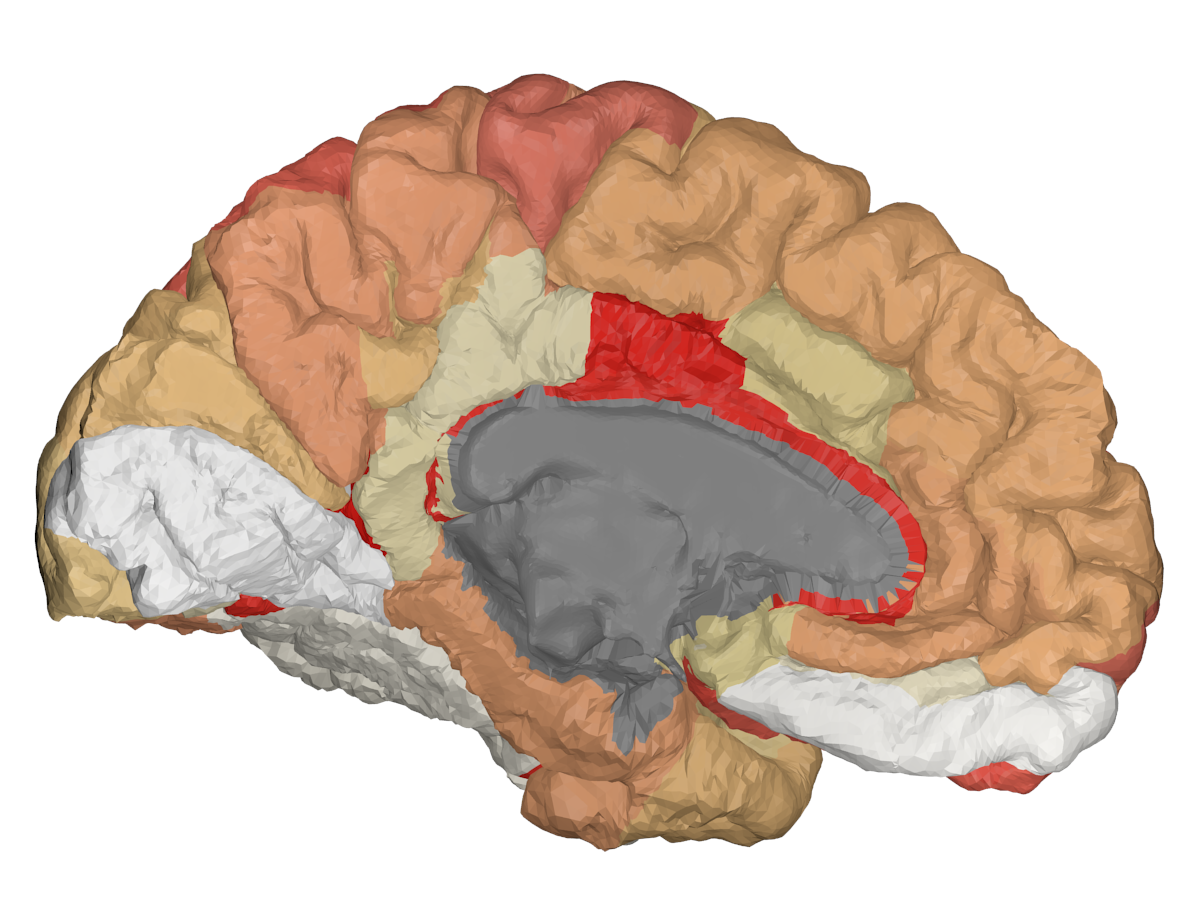} &
    \includegraphics[width=0.32\linewidth]{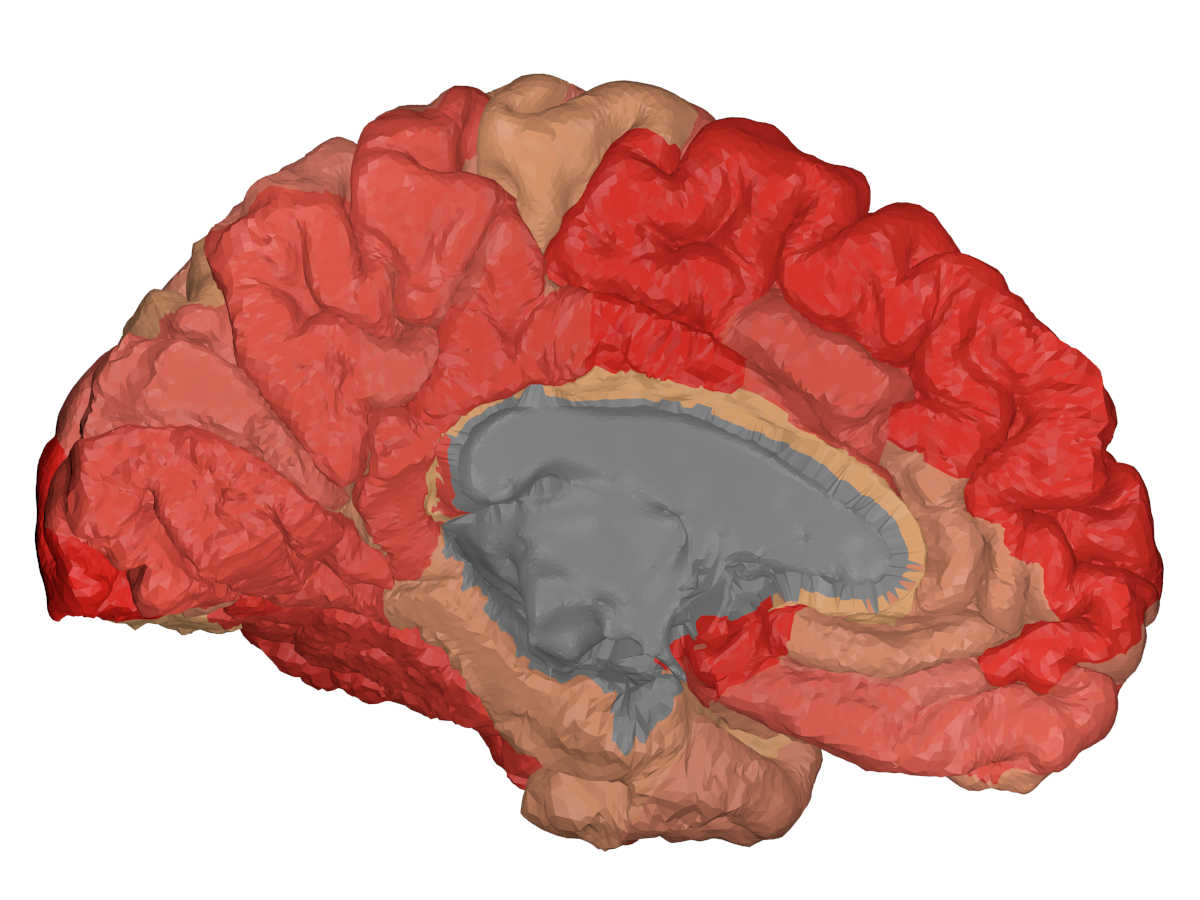} &
    \raisebox{0.26\height}[0pt][0pt]{\includegraphics[width=0.05\textwidth]{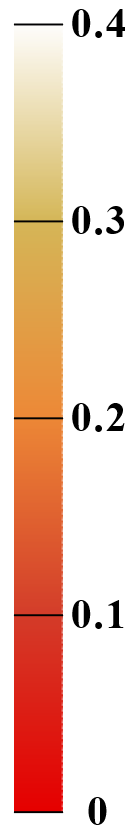}}\\
    \end{tabular}}}
    \vspace{-5pt}
    \caption{\footnotesize Visualization of the averaged absolute Cohen's $d$ between actual distribution and generated distribution on the inner left cortical regions of EMCI subjects. 
    AD-specific regions show better imputation results as 
    lower Cohen's $d$ implies higher correspondence.
    (Row: Source, Column: Target.)
    }
    \vspace{-5pt}
    \label{fig:Cohens}
\end{figure}

The averaged Cohen's $d$ over all ROIs are reported in Table~\ref{tab:Cohens}, 
and their visualization on EMCI subjects are given in Fig.~\ref{fig:Cohens}.
The effect sizes of the data generated by our modality-specific generators were small, i.e., $0.188$ on average, far less than $0.407$ and $0.261$ from individually trained cGAN and WGAN, 
demonstrating the feasibility of our generated data.   

The primary difference between our method and other generative models is the number of generators required for imputation.
Unlike our framework only needs $S$ generators through modality-agnostic embedding,
other conventional generative methods need $S^2$ generators for the same task.
When the sample size is small, training multiple generators 
can be challenging, 
whereas our method 
successfully trains the generators using modality-agnostic embeddings. 

\vspace{-8pt}
\subsection{Qualitative Results}
\label{sec:comparision_quantitative}
\vspace{-5pt}

\begin{figure}[t!]
    \centering
    \renewcommand{\arraystretch}{1.0}
    \renewcommand{\tabcolsep}{0.1cm}
    \small{
    \scalebox{0.58}{
    \begin{tabular}{cccccl}
    &\raisebox{1.1\height}[0pt][0pt]{\textbf{Cortical Thickness}}
    &\raisebox{1\height}[0pt][0pt]{\textbf{Tau}} 
    &\raisebox{1\height}[0pt][0pt]{\textbf{FDG}} 
    &\raisebox{1\height}[0pt][0pt]{\textbf{$\beta$-Amyloid}} 
    & \\
    \raisebox{5\height}[0pt][0pt]{\textbf{True}} &
    \includegraphics[width=0.32\linewidth]{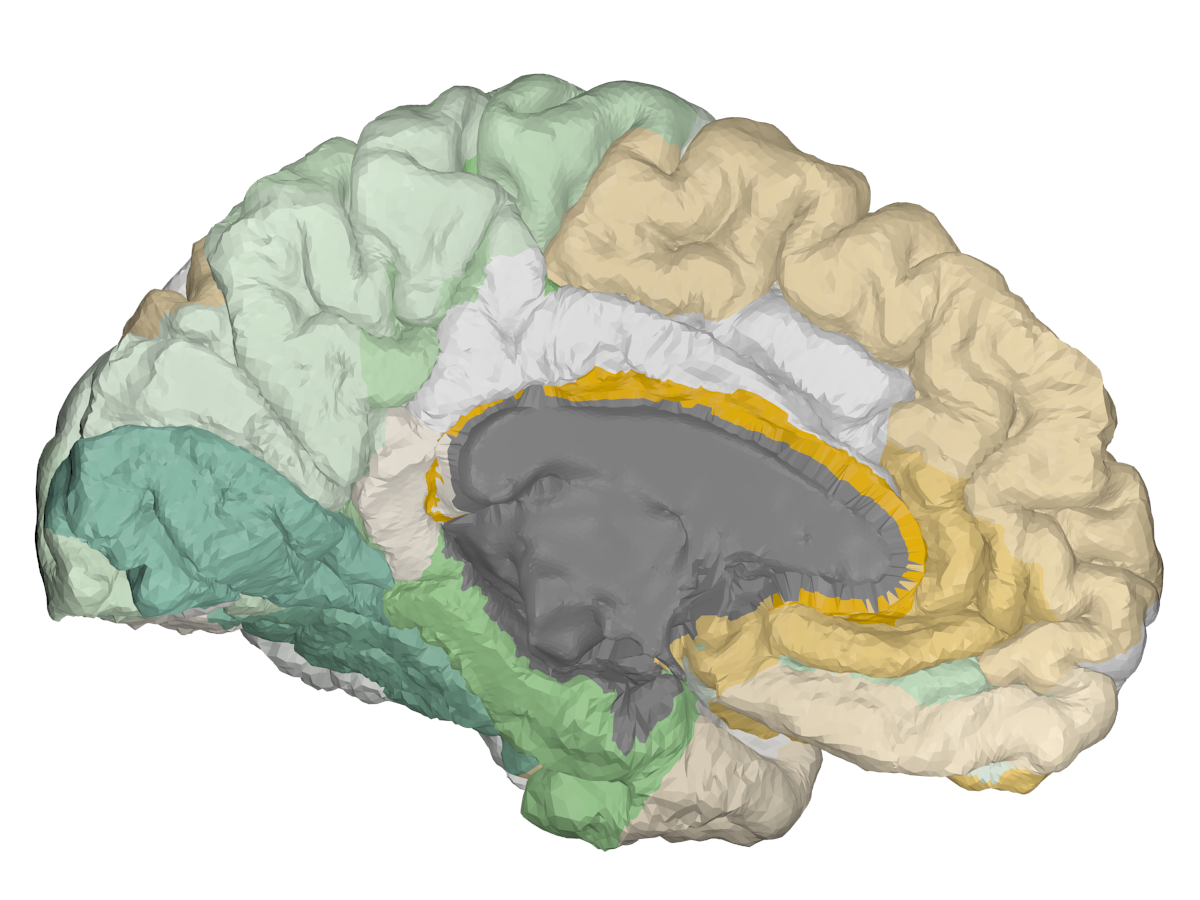}&
    \includegraphics[width=0.32\linewidth]{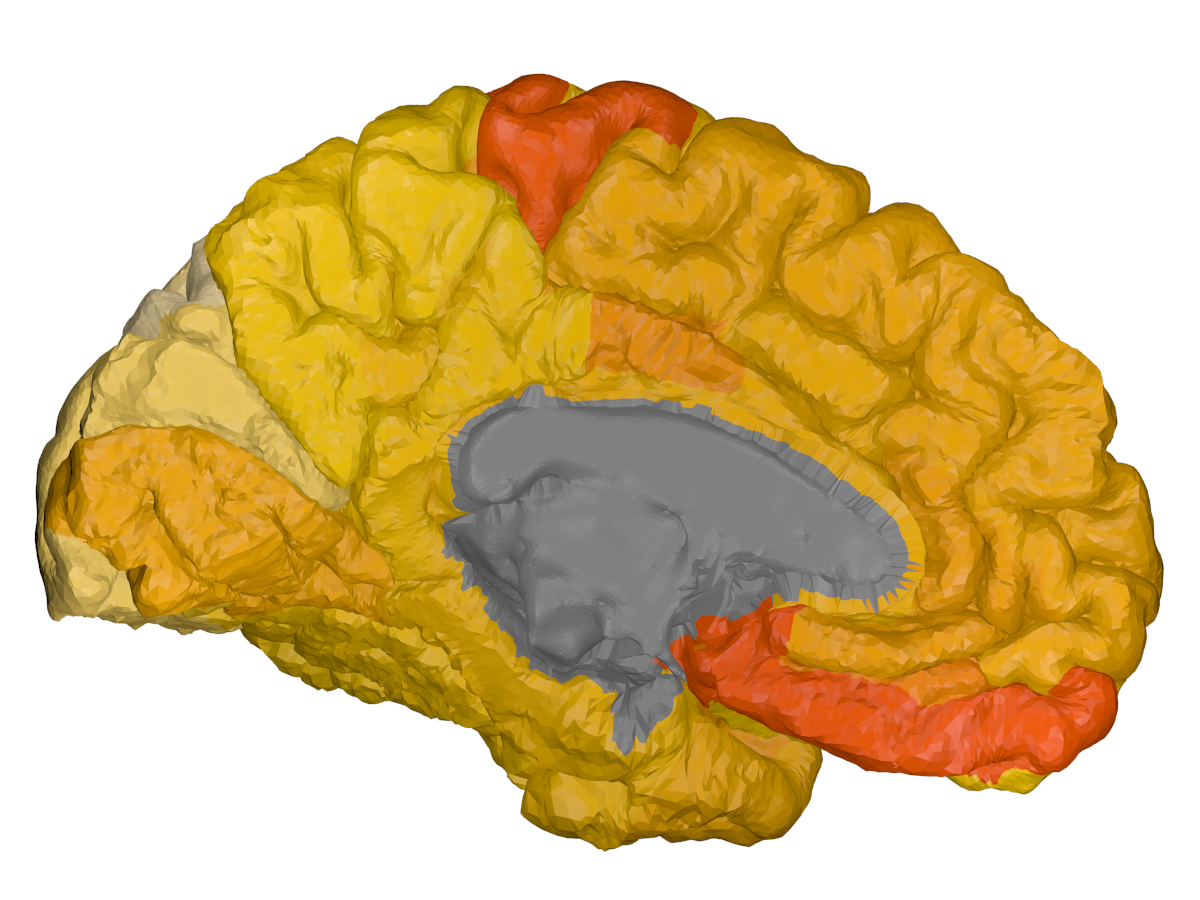} &
    \includegraphics[width=0.32\linewidth]{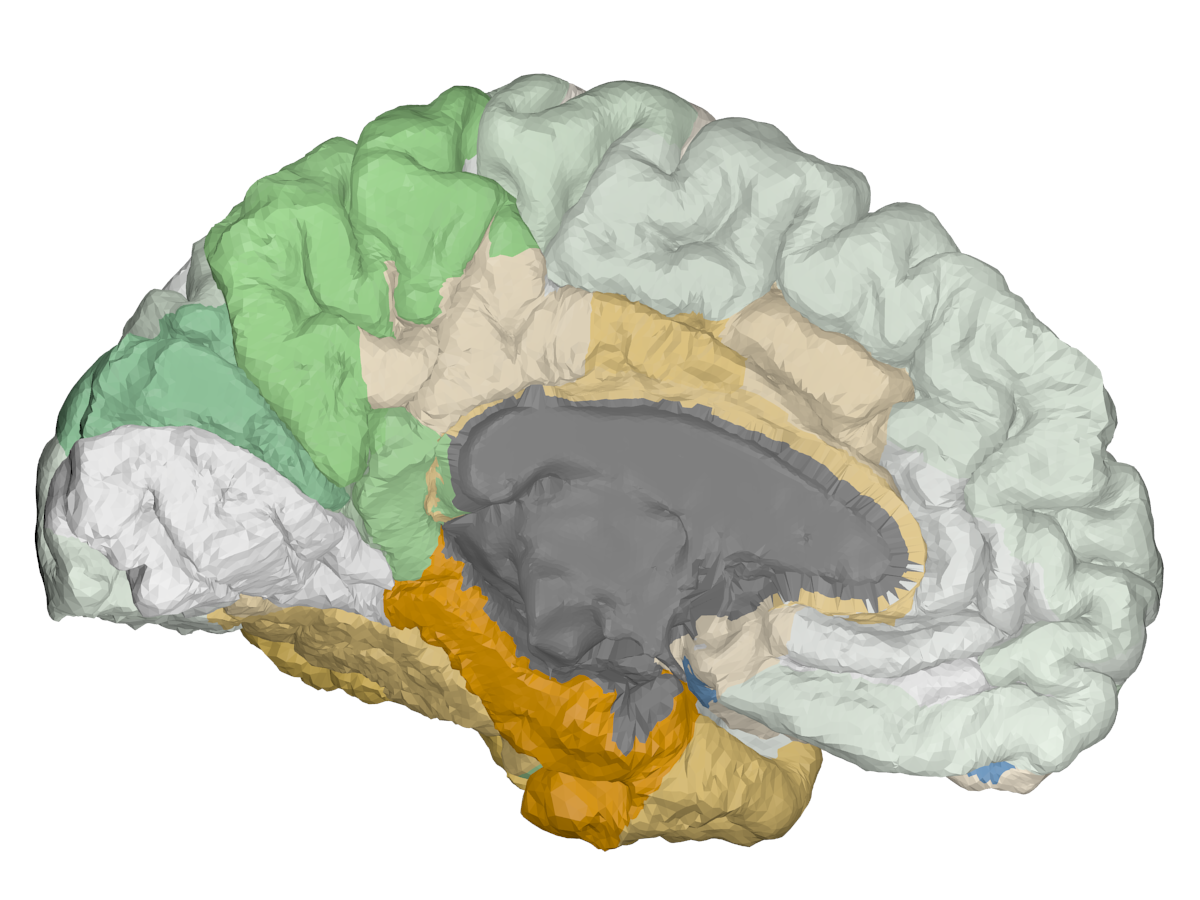} &
    \includegraphics[width=0.32\linewidth]{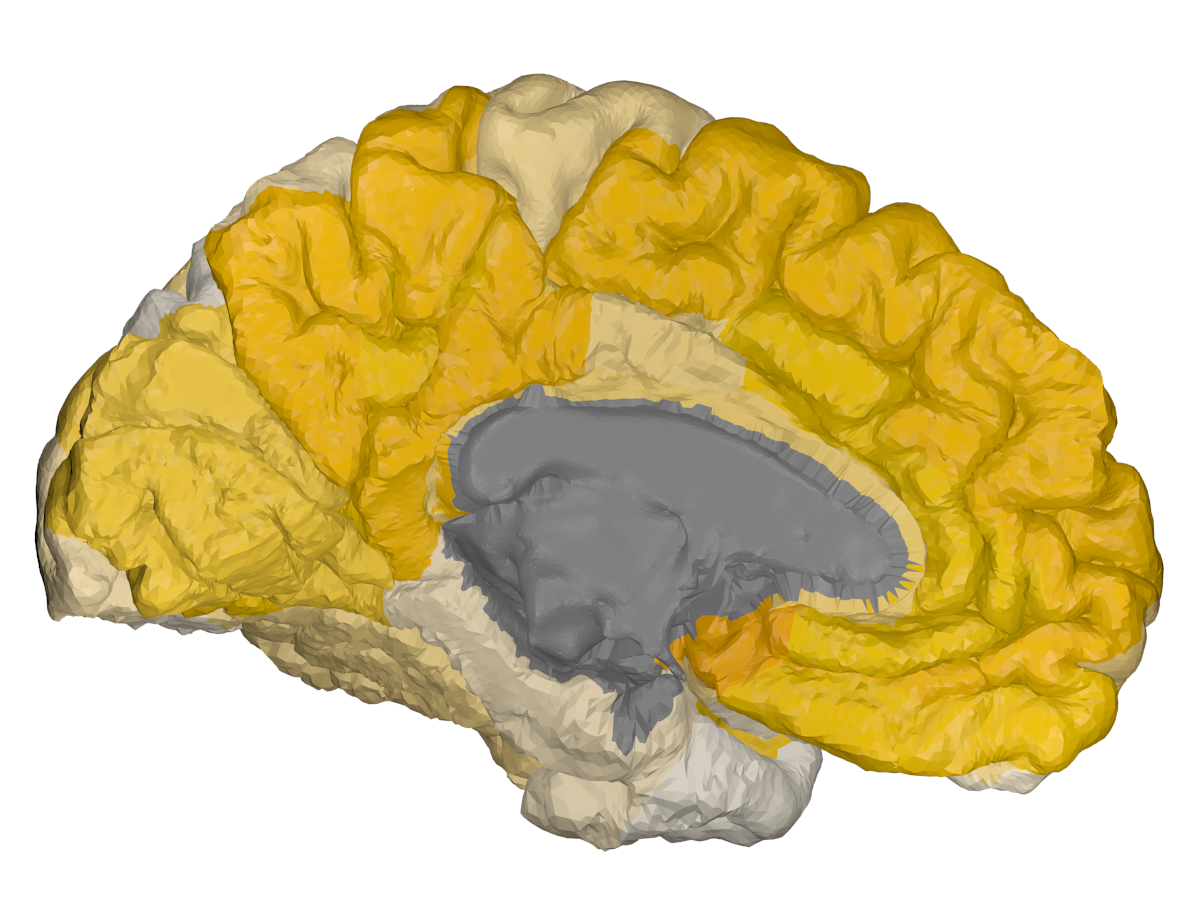} & \\
    \raisebox{5\height}[0pt][0pt]{\textbf{Imputed}} &
    \includegraphics[width=0.32\linewidth]{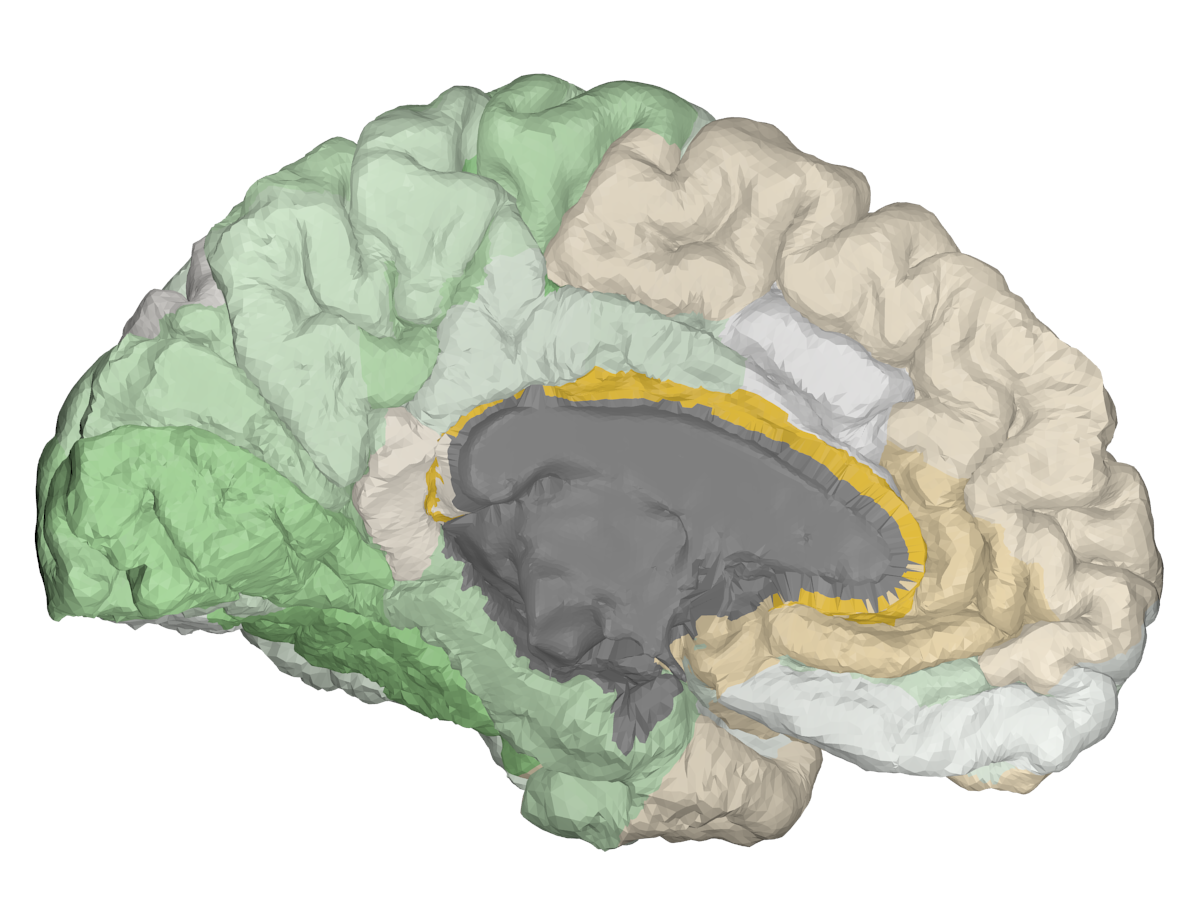}&
    \includegraphics[width=0.32\linewidth]{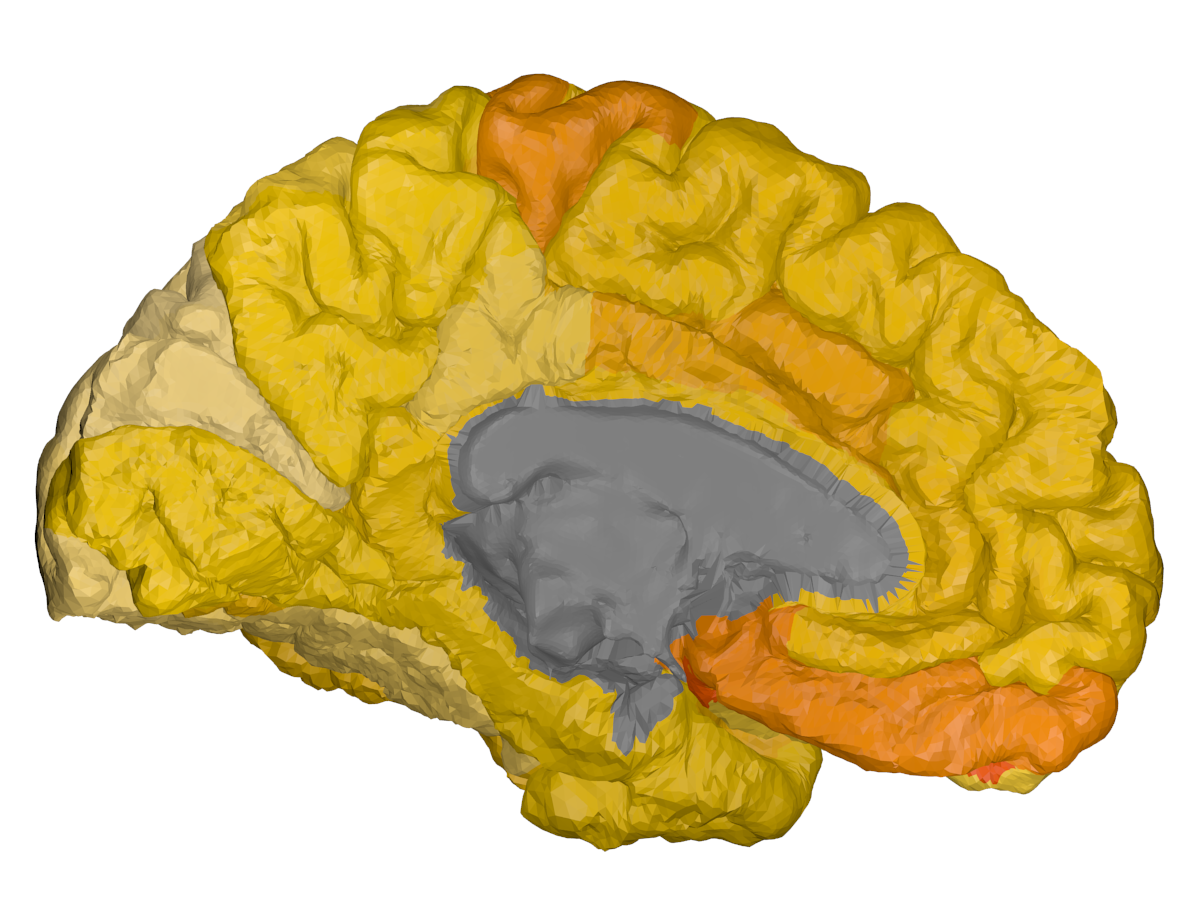} &
    \includegraphics[width=0.32\linewidth]{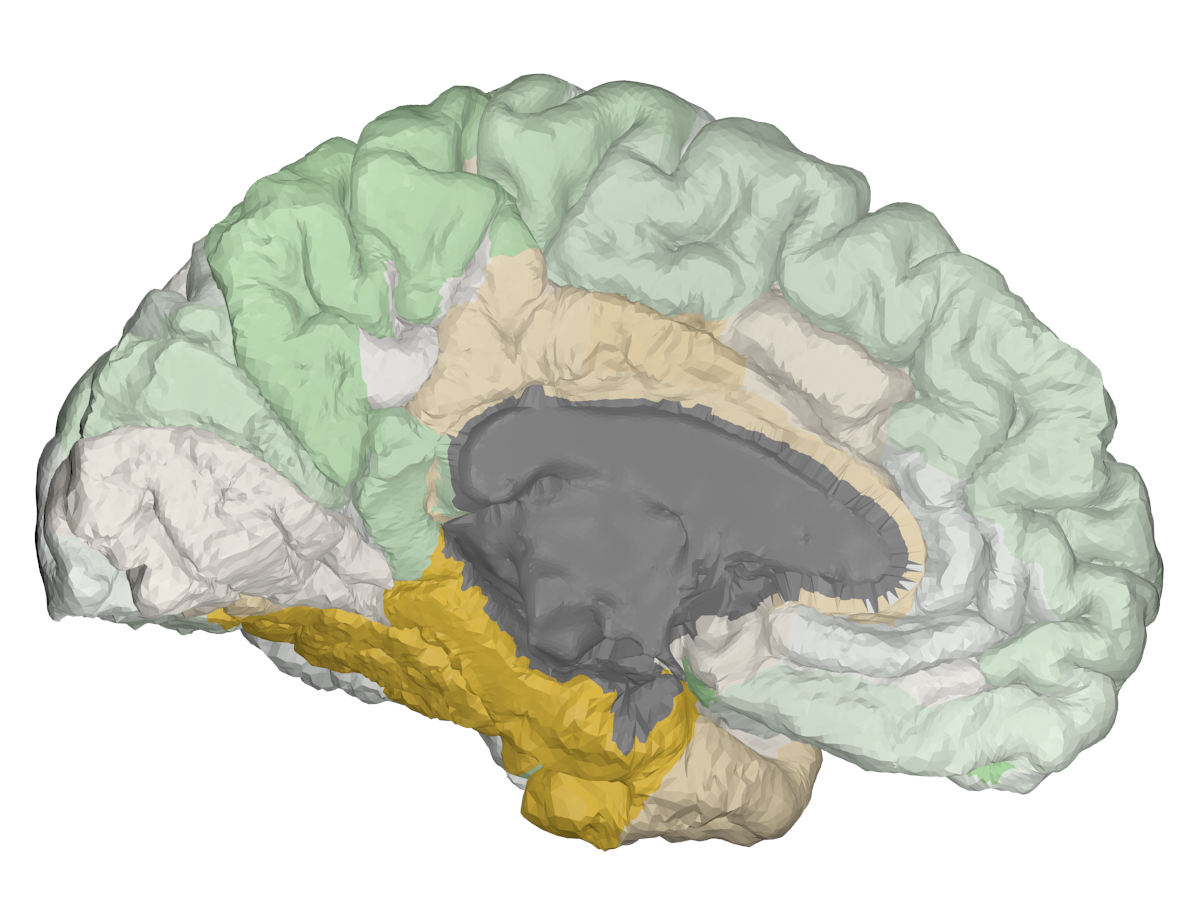} &
    \includegraphics[width=0.32\linewidth]{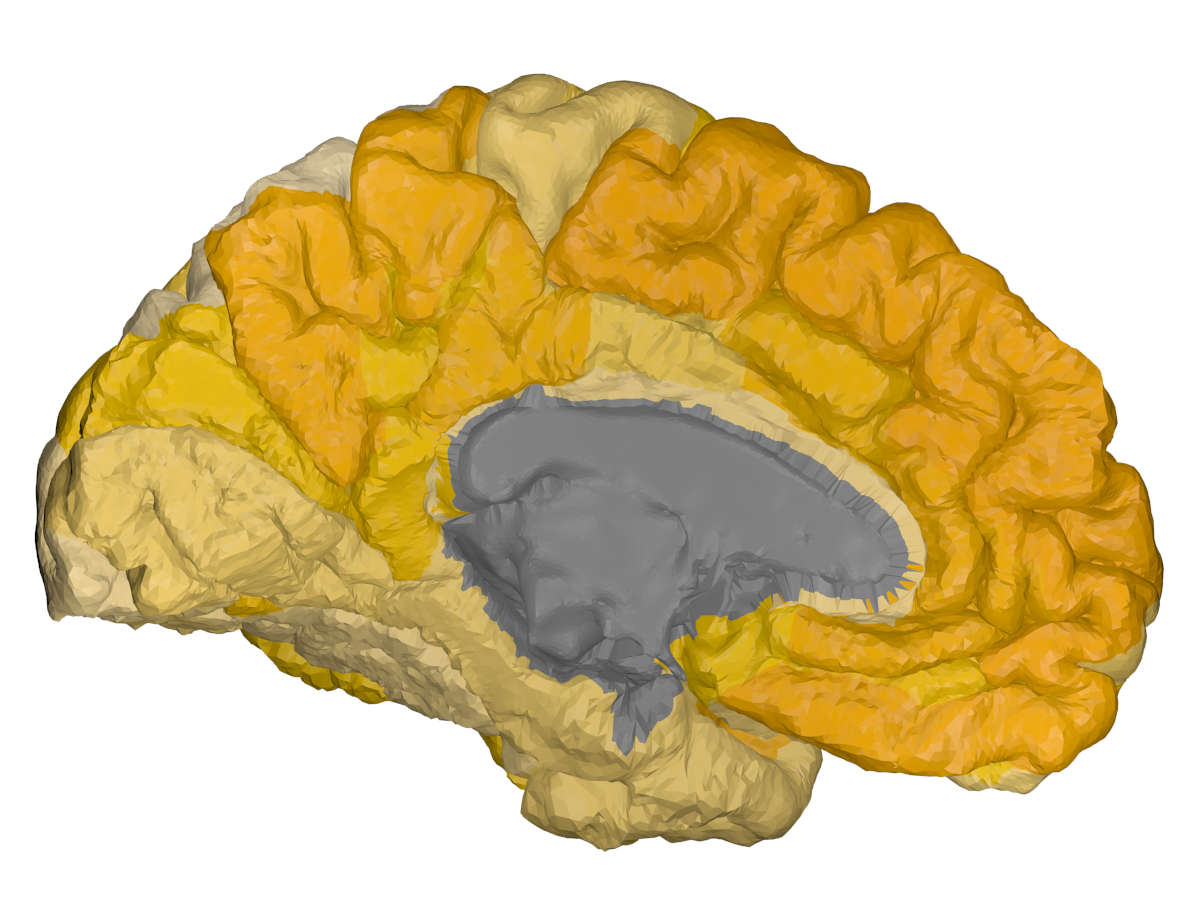} &\\
    \raisebox{5\height}[0pt][0pt]{\textbf{Distance}} &
    \includegraphics[width=0.32\linewidth]{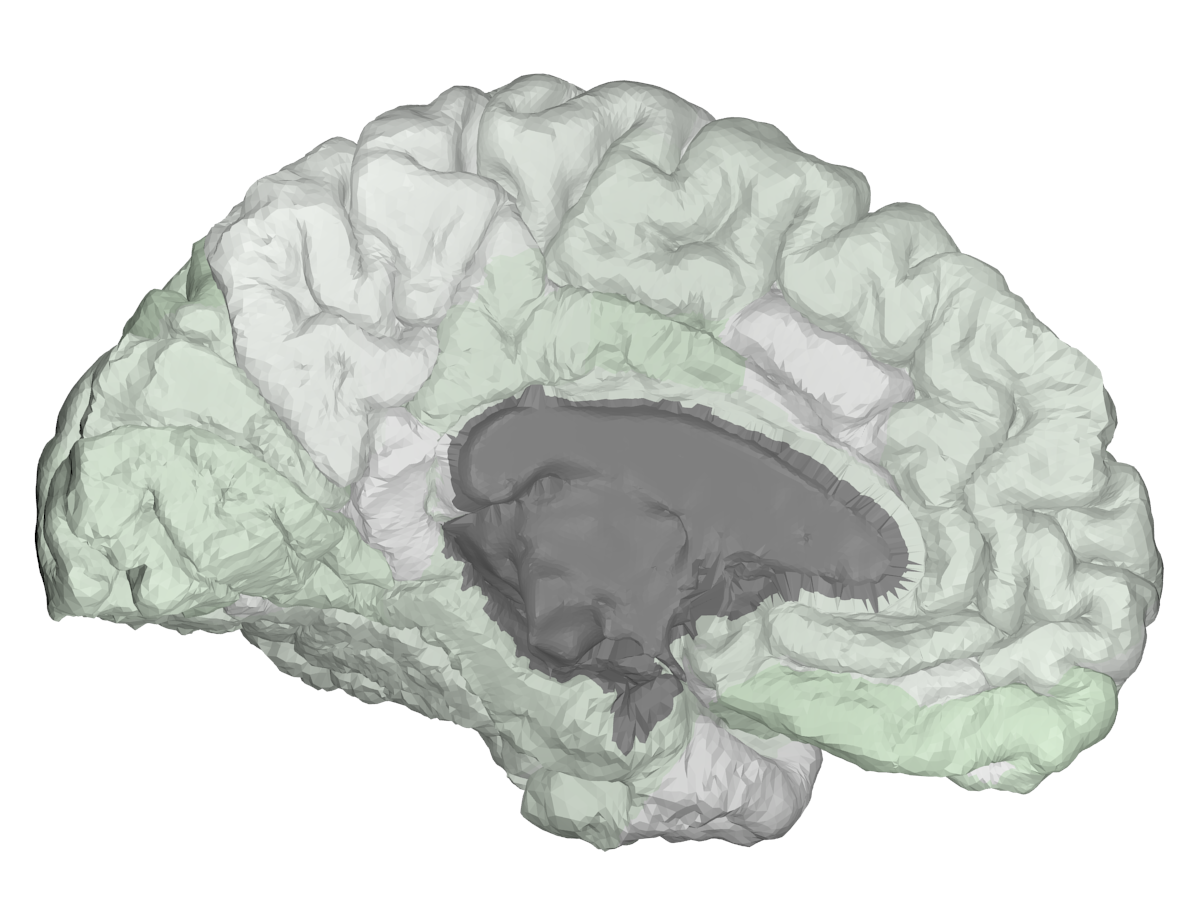}&
    \includegraphics[width=0.32\linewidth]{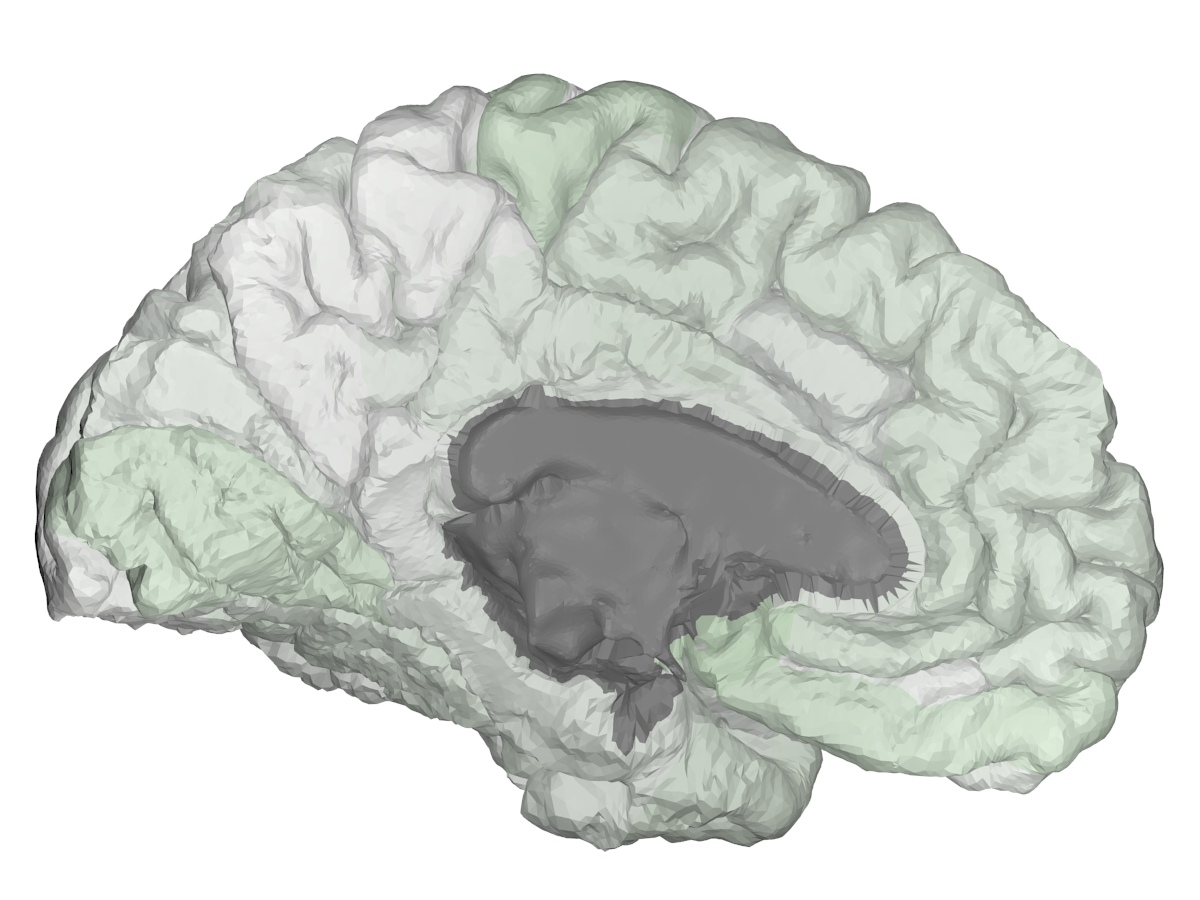} &
    \includegraphics[width=0.32\linewidth]{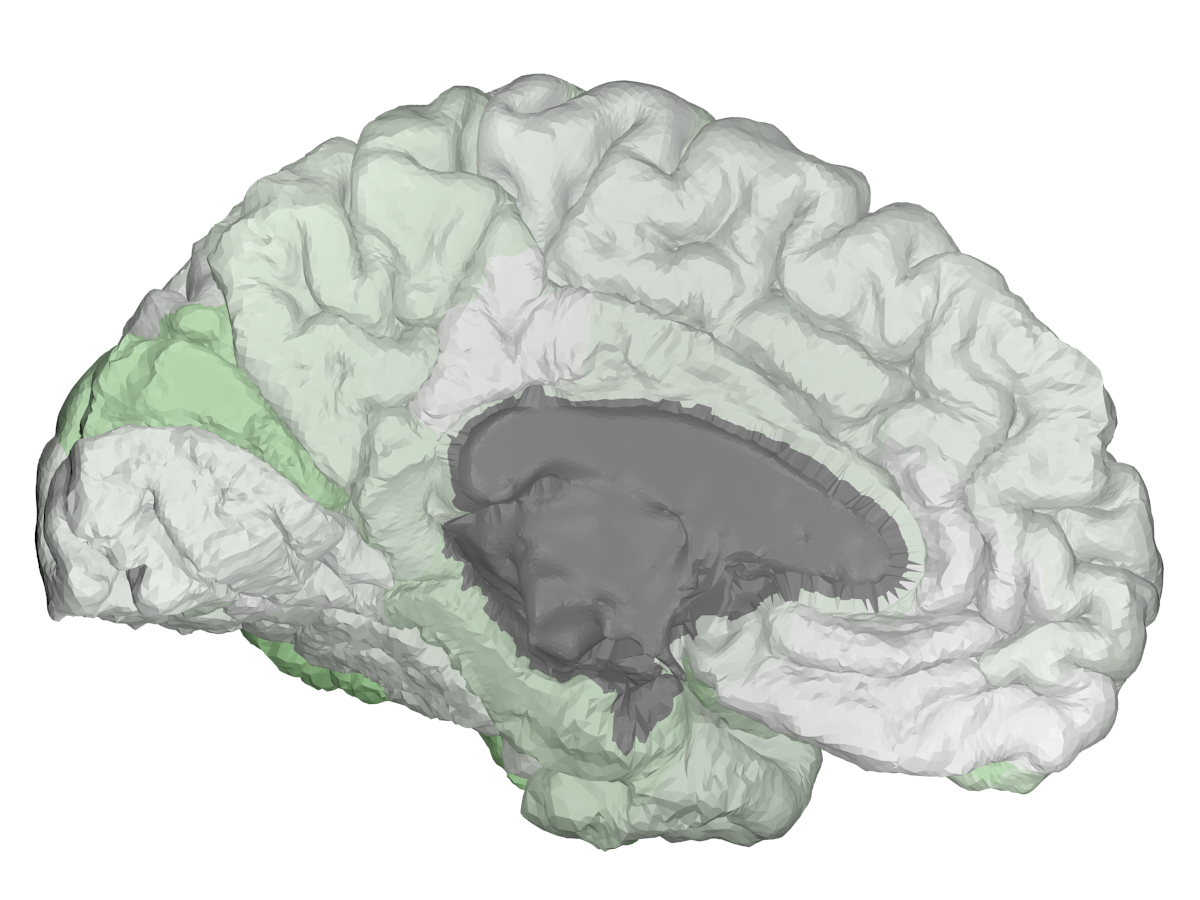} &
    \includegraphics[width=0.32\linewidth]{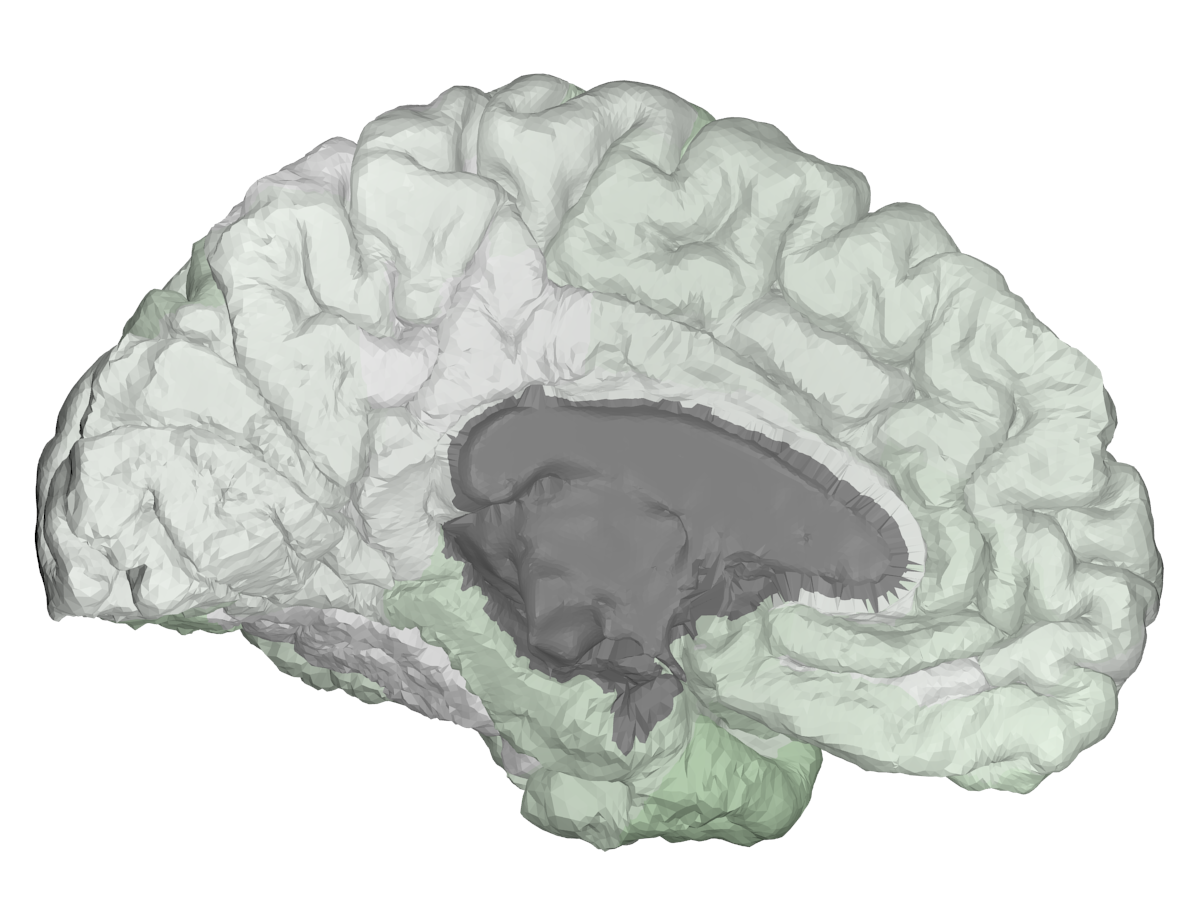} &
    \raisebox{0.17\height}[0pt][0pt]{\includegraphics[width=0.05\textwidth]{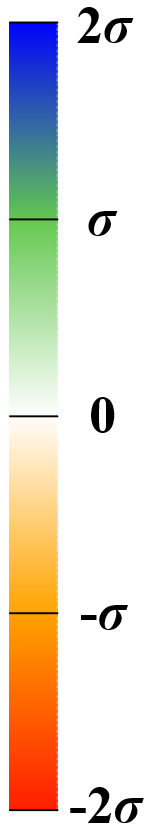}}\\
    \end{tabular}}}
    
    \vspace{-3pt}
    \caption{\footnotesize Visual comparison of 
    observed measurement (Top) and our estimation (Middle) on the inner view of left hemisphere
    from a CN subject. 
    Our estimations were generated from observed CT measurement of the subject.
    All measurements were standardized, and the distance between ground truth and generated result are given at the bottom. 
    BrainPainter~\cite{brainpainter} was used to generate the drawings.
    }
    
    \label{fig:Quantitative_Result}
    \vspace{-5pt}
\end{figure}

 In Fig.~\ref{fig:Quantitative_Result}, we visualized a generated sample (standardized at each ROI), 
 along with the true measurements from a subject in the CN group.  
We selected a subject who underwent all imaging scans and was not used during generator training 
to provide ground truth for our estimation.
From CT of this subject, CT, Tau, FDG and $\beta$-amyloid measures were generated and compared with the ground truth. 
Although we did not train the model with modality pairs from the same subject as they are insufficient,
our model generated every ROI accurately 
toward observed measurement, as visualized in Fig.~\ref{fig:Quantitative_Result} (bottom).

\vspace{-8pt}
\subsection{Downstream MCI Classification Performance}
\label{sec:downstream}
\vspace{-5pt}

In Table~\ref{tab:Qualitative_Result}, 
we reported the performance of the downstream classification with 5-fold CV. 
The MLP using imputed data from our framework outperformed all other baselines 
in accuracy and weighted recall across every MLP classifiers.
Our approach showed an accuracy increase over the baseline in 2-MLP (0.022), 3-MLP (0.026) and 4-MLP (0.040), 
indicating that generated samples enable to train larger models.
However, as we stacked more layers for classifier, 
there was little increase in performance over the baseline with mean imputation, 
indicating that its generated data are limited.  
While usability of mean imputation was limited to downstream classification,  
our primary competitors were cGAN and WGAN as they generate subject-wise estimation. 
Regardless of the model size, 
our model showed superior performance in terms of accuracy and weighted recall over cGAN and WGAN. 
This could be because the two generative baselines were trained using fewer samples
and lacked any precautions against mode-collapse, 
unlike our approach that incorporates content loss.

\vspace{-3pt}
\begin{table}[h!]
\caption{\footnotesize Classification performance (CN/EMCI/LMCI) on ADNI data with all modalities.  
The number of generators required are given in (). 
}
\vspace{-7pt}
\centering
\renewcommand{\arraystretch}{1.25}
\renewcommand{\tabcolsep}{0.28cm}
\scalebox{0.55}{
\begin{tabular}{c|l|c||ccc}
    \toprule[1.5pt]
    \multirow{2}{*}{\textbf{Model}} & \multirow{2}{*}{\textbf{Method}} &  \multirow{2}{*}{\shortstack{\textbf{Imputation}\\\textbf{Concept}}} & \multicolumn{3}{c}{\textbf{Performance}}\\ \cline{4-6}
    & & & \textbf{Accuracy} & \textbf{Precision} &  \textbf{Recall} \\ 
    \hline
    \multirow{5}{*}{\textbf{2-MLP}}& Baseline & No Imputation
    & 0.776 $\pm$ 0.032  & 0.794 $\pm$ 0.030 & 0.776 $\pm$ 0.032 \\
    & Mean & Representative Value
    & 0.792 $\pm$ 0.081 & 0.806 $\pm$ 0.052 & 0.792 $\pm$ 0.081 \\ 
    & cGAN~\cite{cGAN} & Generative Model (16)
    & 0.784 $\pm$ 0.092 & 0.743 $\pm$ 0.183 & 0.784 $\pm$ 0.092 \\ 
    & WGAN~\cite{WGAN} & Generative Model (16)
    & 0.797 $\pm$ 0.075 & \textbf{0.834} $\pm$ 0.072 & 0.797 $\pm$ 0.075 \\ \cline{2-6}
    & \textbf{Ours} & Generative Model (4)
    & \textbf{0.798} $\pm$ 0.040 & 0.789 $\pm$ 0.048 & \textbf{0.798} $\pm$ 0.040 \\ \hline

    \multirow{5}{*}{\textbf{3-MLP}}& Baseline & No Imputation
    & 0.785 $\pm$ 0.039  & \textbf{0.811} $\pm$ 0.032 & 0.785 $\pm$ 0.039 \\
    & Mean & Representative Value
    & 0.802 $\pm$ 0.096 & 0.790 $\pm$ 0.087 & 0.802 $\pm$ 0.096 \\ 
    & cGAN~\cite{cGAN} & Generative Model (16)
    & 0.805 $\pm$ 0.076 & 0.809 $\pm$ 0.095 & 0.805 $\pm$ 0.076 \\ 
    & WGAN~\cite{WGAN} & Generative Model (16)
    & 0.784 $\pm$ 0.072 & 0.770 $\pm$ 0.093 & 0.784 $\pm$ 0.072 \\ \cline{2-6}
    & \textbf{Ours} & Generative Model (4)
    & \textbf{0.811} $\pm$ 0.061 & 0.791 $\pm$ 0.076 & \textbf{0.811} $\pm$ 0.061 \\\hline

    \multirow{5}{*}{\textbf{4-MLP}}& Baseline & No Imputation
    & 0.785 $\pm$ 0.018  & \textbf{0.804} $\pm$ 0.018 & 0.785 $\pm$ 0.018 \\
    & Mean & Representative Value
    & 0.788 $\pm$ 0.084 & 0.771 $\pm$ 0.092 & 0.788 $\pm$ 0.084 \\ 
    & cGAN~\cite{cGAN} & Generative Model (16)
    & 0.788 $\pm$ 0.105 & 0.766 $\pm$ 0.105 & 0.788 $\pm$ 0.105 \\ 
    & WGAN~\cite{WGAN} & Generative Model (16)
    & 0.792 $\pm$ 0.054 & 0.796 $\pm$ 0.093 & 0.792 $\pm$ 0.056 \\ \cline{2-6}
    & \textbf{Ours} & Generative Model (4)
    & \textbf{0.825} $\pm$ 0.018 & 0.793 $\pm$ 0.051 & \textbf{0.825} $\pm$ 0.018 \\
    \bottomrule[1.5pt]
    \end{tabular}}
\label{tab:Qualitative_Result}
\vspace{-10pt}
\end{table}
\vspace{-8pt}
\section{Conclusion}
\label{sec:conclusion}
\vspace{-5pt}

In this paper, we propose a novel framework that generates unobserved imaging measures for specific subjects using their existing measures.
To reduce the need for taking several imaging scans,  
our framework addresses the imputation of missing measures 
by transferring modality-specific style while preserving AD-specific content.
Experimental results on the ADNI study show that our model provides a probable estimation of target modality for individual subjects,
which yields similar distributions of generated measurements to those from observed data and helps downstream analyses. 
Since our work is applicable regardless of modality type, 
our approach has the potential to be adopted by other neuroimaging studies that are limited by missing measures. 

\section{Acknowledgement}
\vspace{-2mm}
\sloppy This research was supported by NRF-2022R1A2C2092336 (50\%), IITP-2022-0-00290 (20\%), IITP-2019-0-01906 (AI Graduate Program at POSTECH, 10\%) funded by MSIT, HU22C0171 (10\%) and HU22C0168 (10\%) funded by MOHW from South Korea.

\vspace{-2mm}
\bibliographystyle{IEEEbib}
\bibliography{refs}

\begin{thebibliography}{10}

\bibitem{baek2023}
Seunghun Baek, Injun Choi, et~al.,
\newblock ``Learning covariance-based multi-scale representation of
  neuroimaging measures for alzheimer classification,''
\newblock in {\em ISBI}. IEEE, 2023, pp. 1--5.

\bibitem{sim2024}
Jaeyoon Sim, Sooyeon Jeon, InJun Choi, Guorong Wu, and Won~Hwa Kim,
\newblock ``Learning to approximate adaptive kernel convolution on graphs,''
\newblock in {\em Proceedings of the AAAI Conference on Artificial
  Intelligence}, 2024, vol.~38, pp. 4882--4890.

\bibitem{correlation1}
Rik Ossenkoppele, Smith, et~al.,
\newblock ``Associations between tau, a$\beta$, and cortical thickness with
  cognition in alzheimer disease,''
\newblock {\em Neurology}, vol. 92, no. 6, pp. e601--e612, 2019.

\bibitem{correlation2}
Theresa~M Harrison, Du, et~al.,
\newblock ``Distinct effects of beta-amyloid and tau on cortical thickness in
  cognitively healthy older adults,''
\newblock {\em Alzheimer's \& dementia}, vol. 17, no. 7, pp. 1085--1096, 2021.

\bibitem{correlation3}
Anna Rubinski, Nicolai Franzmeier, Julia Neitzel, Michael Ewers, and
  Alzheimer’s Disease Neuroimaging~Initiative (ADNI),
\newblock ``Fdg-pet hypermetabolism is associated with higher tau-pet in mild
  cognitive impairment at low amyloid-pet levels,''
\newblock {\em Alzheimer's Research \& Therapy}, vol. 12, pp. 1--12, 2020.

\bibitem{correlation4}
Jennifer~L Whitwell, Graff-Radford, et~al.,
\newblock ``Imaging correlations of tau, amyloid, metabolism, and atrophy in
  typical and atypical alzheimer's disease,''
\newblock {\em Alzheimer's \& Dementia}, vol. 14, no. 8, pp. 1005--1014, 2018.

\bibitem{MRI2PET1}
Jin Zhang, Xiaohai He, Linbo Qing, Feng Gao, and Bin Wang,
\newblock ``Bpgan: Brain pet synthesis from mri using generative adversarial
  network for multi-modal alzheimer’s disease diagnosis,''
\newblock {\em Computer Methods and Programs in Biomedicine}, vol. 217, pp.
  106676, 2022.

\bibitem{MRI2PET2}
Yongsheng Pan, Mingxia Liu, Chunfeng Lian, Tao Zhou, Yong Xia, and Dinggang
  Shen,
\newblock ``Synthesizing missing pet from mri with cycle-consistent generative
  adversarial networks for alzheimer’s disease diagnosis,''
\newblock in {\em MICCAI}. Springer, 2018, pp. 455--463.

\bibitem{cycleGAN}
Jun-Yan Zhu, Taesung Park, Phillip Isola, and Alexei~A Efros,
\newblock ``Unpaired image-to-image translation using cycle-consistent
  adversarial networks,''
\newblock in {\em ICCV}, 2017, pp. 2223--2232.

\bibitem{cGAN}
Mehdi Mirza and Simon Osindero,
\newblock ``Conditional generative adversarial nets,''
\newblock 2014.

\bibitem{WGAN}
Martin Arjovsky, Soumith Chintala, and L{\'e}on Bottou,
\newblock ``Wasserstein generative adversarial networks,''
\newblock in {\em International conference on machine learning}. PMLR, 2017,
  pp. 214--223.

\bibitem{donders2006gentle}
A~Rogier~T Donders, Geert~JMG Van Der~Heijden, Theo Stijnen, and Karel~GM
  Moons,
\newblock ``A gentle introduction to imputation of missing values,''
\newblock {\em Journal of clinical epidemiology}, vol. 59, no. 10, pp.
  1087--1091, 2006.

\bibitem{domain_adversarial}
Yaroslav Ganin, Ustinova, et~al.,
\newblock ``Domain-adversarial training of neural networks,''
\newblock {\em JMLR}, vol. 17, no. 1, pp. 2096--2030, 2016.

\bibitem{style_transfer1}
Leon~A Gatys, Alexander~S Ecker, and Matthias Bethge,
\newblock ``Image style transfer using convolutional neural networks,''
\newblock in {\em CVPR}, 2016, pp. 2414--2423.

\bibitem{style_transfer2}
Fujun Luan, Sylvain Paris, Eli Shechtman, and Kavita Bala,
\newblock ``Deep photo style transfer,''
\newblock in {\em CVPR}, 2017, pp. 4990--4998.

\bibitem{jack2008alzheimer}
Clifford~R Jack~Jr, Bernstein, et~al.,
\newblock ``The alzheimer's disease neuroimaging initiative (adni): Mri
  methods,''
\newblock {\em Journal of Magnetic Resonance Imaging: An Official Journal of
  the International Society for Magnetic Resonance in Medicine}, vol. 27, no.
  4, pp. 685--691, 2008.

\bibitem{destrieux2010automatic}
Christophe Destrieux, Bruce Fischl, Anders Dale, and Eric Halgren,
\newblock ``Automatic parcellation of human cortical gyri and sulci using
  standard anatomical nomenclature,''
\newblock {\em Neuroimage}, vol. 53, no. 1, pp. 1--15, 2010.

\bibitem{thie2004understanding}
Joseph~A Thie,
\newblock ``Understanding the standardized uptake value, its methods, and
  implications for usage,''
\newblock {\em Journal of Nuclear Medicine}, vol. 45, no. 9, pp. 1431--1434,
  2004.

\bibitem{rapoport2000role}
Mark Rapoport, Robert van Reekum, and Helen Mayberg,
\newblock ``The role of the cerebellum in cognition and behavior: a selective
  review,''
\newblock {\em The Journal of neuropsychiatry and clinical neurosciences}, vol.
  12, no. 2, pp. 193--198, 2000.

\bibitem{AdamW}
Ilya Loshchilov and Frank Hutter,
\newblock ``Decoupled weight decay regularization,''
\newblock 2017.

\bibitem{rice2005comparing}
Marnie~E Rice and Grant~T Harris,
\newblock ``Comparing effect sizes in follow-up studies: Roc area, cohen's d,
  and r,''
\newblock {\em Law and human behavior}, vol. 29, pp. 615--620, 2005.

\bibitem{brainpainter}
R{\u{a}}zvan~V Marinescu et~al.,
\newblock ``Brainpainter: A software for the visualisation of brain structures,
  biomarkers and associated pathological processes,''
\newblock in {\em Multimodal Brain Image Analysis and Mathematical Foundations
  of Computational Anatomy: 4th International Workshop, MBIA}. Springer, 2019,
  pp. 112--120.

\end{thebibliography}

\end{document}